\documentclass[abstractpage,ackpage,copyright,figures,tables,phd]{Prelude/uithesis}

\usepackage{amsfonts}
\usepackage{amsmath}
\usepackage{amssymb}
\usepackage{array}
\usepackage{float}
\usepackage[toc,page,titletoc]{appendix}
\usepackage{longtable}
\usepackage{graphicx}
\usepackage[normalem]{ulem}
\usepackage{nicefrac}
\usepackage{units}
\usepackage{rotating}
\usepackage{afterpage}
\usepackage{caption}

\usepackage{mathtools}
\usepackage{mathrsfs}
\usepackage{tensor}
\usepackage{cite}
\usepackage{latexsym}
\usepackage{extarrows}
\usepackage{epsfig}
\usepackage{subfig}
\usepackage{graphics}
\usepackage{dsfont}

\usepackage{etoolbox}
\apptocmd{\thebibliography}{\interlinepenalty 10000\relax}{}{}

\usepackage{enumitem}
\setlist{noitemsep,topsep=1pt, partopsep=4pt, parsep=2pt}
\setenumerate{noitemsep,topsep=1pt, partopsep=4pt, parsep=2pt}

\let\savedCaption=\caption
\renewcommand*{\caption}[2][\shortcaption]{%
  \def\shortcaption{#2}
  \normalsize\singlespace
  \savedCaption[#1]{#2}}

\newcommand{\al}[1]{\begin{align}#1\end{align}}
\newcommand{\bs}{\begin{split}}
\newcommand{\es}{\end{split}}
\newcommand{\eqr}{\eqref}
\newcommand{\mc}{\mathcal}
\newcommand{\pa}[1]{\left(#1\right)}
\newcommand{\pb}[1]{\left[#1\right]}

\newcommand{\expval}[1]{\langle #1 \rangle}

\title{TOPOLOGICALLY MASSIVE YANG-MILLS THEORY \\AND LINK INVARIANTS}

\author{Tuna Yildirim}
\dept{Physics}
\setboolean{multipleSupervisors}{false}
\advisor{Professor Vincent G. J. Rodgers}
\memberOne{Vincent G. J. Rodgers}
\memberTwo{Charles Frohman}
\memberThree{Craig E. Pryor}
\memberFour{Wayne N. Polyzou}
\memberFive{Mary Hall Reno}
\submitdate{December 2014}
\copyrightyear{2014}

\setboolean{copyrightpage}{true}

\dedication{Prelude/dedication}
\setboolean{dedicationpage}{true}

\ackfile{Prelude/thesisAck}
\abstractfile{Prelude/thesisAbstract}
\publicabstractfile{Prelude/publicAbstract}

\setboolean{figures}{true}
\setboolean{tables}{false}

\begin{document}
\frontmatter

\chapter{Introduction}\label{ch:intro}
\addtocontents{toc}{\protect\vspace{-\li}}
 
Chern-Simons theory has been extensively studied and is a very important part of mathematical physics, mostly because of its connection with the link invariants of knot theory. This was first demonstrated by Witten \cite{Witten:1988hf} using 2D conformal field theories related to Chern-Simons theory. Witten showed that Wilson loop expectation values of Chern-Simons theory are given by link invariant polynomials which can be recursively calculated from skein relations. Later, Cotta-Ramusino et al. \cite{CottaRamusino:1989rf} derived skein relations for Chern-Simons theory using only 3D field theory techniques. No matter what method is used, one crucial requirement for relating Wilson loop expectation values to link invariants is that the theory must be topological; in other words, the action must be metric free. Thus, it is not possible to do this with a Yang-Mills action because of its metric dependence. The main goal of this thesis is to find out how Chern-Simons link invariants are modified with the presence of a Yang-Mills term at large but finite distances, where the metric contribution is very small, hence the theory is almost topological.

It is well known that adding a Chern-Simons term to Yang-Mills action in 2+1 dimensions gives mass to gauge bosons\cite{schonfeld1981mass}. A considerable amount of work on topologically massive Yang-Mills theory was done by Deser, Jackiw and Templeton\cite{Deser1982372,Deser2,Deser1984371} in the early 80s and some of the later work can be found in refs. \citen{Gonzales1986104, horvathy, Evens, martinez1989constrained, Giavarini1992222, Giavarini1993qd, Asorey1993477, Grignani1997360, Karabali:1999ef, dunne2000magnetic, Canfora2013}. Although topologically massive Yang-Mills theory is more complicated than Chern-Simons theory, there are surprising similarities between these theories, partly because conjugate momenta of both theories are given by gauge fields. An interesting example that shows the similarity of these two theories is the classical equivalence, first observed by Lemes et al\cite{Lemes:1997vx,Lemes:1998md}. This equivalence shows that classically it is possible to write the topologically massive Yang-Mills action as a pure Chern-Simons action via a non-linear redefinition of the gauge fields. One would expect that it might be possible to extend this equivalence to quantum level, but we will show that phase space geometry does not allow this. Instead, we will obtain a more complicated equivalence between the observables of both theories at large but finite distances.

Since pure Yang-Mills theory in 2+1 dimensions has a mass gap\cite{Karabali:1996iu}, the theory is trivial at very large distances. However, in topologically massive Yang-Mills theory that is not the case. The mass gap of this theory is proportional to the topological mass $m$\cite{Karabali:1999ef}. Studying large values of topological mass is equivalent to scaling up the metric or looking at large distances. In this thesis, we study the theory at large but finite distances by neglecting the second and higher order terms in $1/m$, while keeping the first. This leads to an almost topological theory with a small contribution from Yang-Mills, which allows us to write topologically massive Yang-Mills theory observables in terms of Wilson loop expectation values of Chern-Simons theory. This means that, not only in the pure Chern-Simons limit but also in the \emph{near} Chern-Simons limit, one only needs skein relations to calculate topologically massive Yang-Mills theory observables. We also study pure Yang-Mills theory at the same distance scale and obtain similar interesting results.

The thesis is organized as follows. Most of the necessary background knowledge is given in \autoref{ch:bckgrd}, including introductions to knot theory (\autoref{sec:knot}) and geometric quantization (\autoref{sec:geoquan}). In \autoref{ch:cs}, the methods described in \autoref{ch:bckgrd} are applied to Chern-Simons theory. In \autoref{ch:tmym}, similar methods are applied on a more complicated theory: topologically massive Yang-Mills. In \autoref{ch:ym}, we look at large distance behavior of pure Yang-Mils theory. Finally, in \autoref{ch:conc}, the results obtained in \autoref{ch:tmym} and \autoref{ch:ym} are discussed.
\chapter{Background}\label{ch:bckgrd}

\section{Yang-Mills Theory}

Yang-Mills Lagrangian is the non-abelian generalization of the Maxwell Lagrangian, written as
\al{
\label{eq:YMlagr}
\mc{L}=\frac{1}{2e^2}Tr(F_{\mu\nu}F^{\mu\nu}),
}
where the constant $e^2$ is dimensionless in 4D and has the dimension of mass in 3D. The curvature $F_{\mu\nu}$ is given by the commutator of the covariant derivatives, as $F_{\mu\nu}=[D_\mu, D_\nu]$, where $D_\mu=\partial_\mu+A_\mu$. $A_\mu$ can be expanded as $A_\mu=-i A_\mu^a t^a$, where $t^a$ are matrix representatives of the generators of Lie algebra $[t^a, t^b]=if^{abc}t^c$. Here, $t^a$ are hermitian matrices and $f^{abc}$ are real valued structure constants. For $G=SU(N)$ where $G$ is the gauge group, $a,b,c=1,2,\dotsc,N^2-1$. In the fundamental representation, the trace is normalized as $Tr(t^a t^b)=\frac{1}{2}\delta^{ab}$. Expanding the trace, the Lagrangian \eqref{eq:YMlagr} can be rewritten as
\al{
\mc{L}=-\frac{1}{4e^2}F^a_{\mu\nu}F^{a\mu\nu}.
} 
Under a gauge transformation $A_\mu \rightarrow A_\mu^g = g A_\mu g^{-1} -\partial_\mu g g^{-1}$, $F_{\mu\nu}$ transforms covariantly, as
\al{
F_{\mu\nu}(A) \rightarrow F_{\mu\nu}(A^g)= g F_{\mu\nu}(A) g^{-1},
}
where $g \in G$. The cyclicity property of the trace makes the Lagrangian \eqref{eq:YMlagr} gauge invariant. The equations of motion are given by
\al{
D_\mu F^{\mu\nu}=0.
}
Covariant derivatives satisfy the Jacobi identity
\al{
[D_\mu,[D_\nu,D_\alpha]]+[D_\nu,[D_\alpha,D_\mu]]+[D_\alpha,[D_\mu,D_\nu]]=0,
}
which leads to the Bianchi identity
\al{
D_\alpha F_{\mu\nu} + D_\mu F_{\nu\alpha} + D_\nu F_{\alpha\mu}=0.
}

There is another important gauge invariant Lagrangian that can be constructed using $F_{\mu\nu}$. It is called the Pontryagin density and it is given by
\al{
\label{eq:pontr}
\mc{L}_P=-\frac{1}{32\pi^2} \epsilon^{\mu\nu\alpha\beta} Tr(F_{\mu\nu} F_{\alpha\beta}).
}
This Lagrangian can be written as a total derivative and it is related to the Chern-Simons Lagrangian as we will see in \autoref{sec:csint}.

Non-abelian Yang-Mills theory has been very successful in explaining weak and strong interactions. Thus, it is a very important part of the standard model.

\section{Problems of Quantum Chromodynamics}

The widely accepted model for Quantum Chromo-Dynamics(QCD) is an SU(3) gauge theory given by a Yang-Mills Lagrangian for gluons, plus a Dirac Lagrangian for quarks. An additional small topological term is also a possibility, to explain CP violation. This term is given by \eqref{eq:pontr}.

The correct QCD action is expected to have a few important features. One of these is called the ``asymptotic freedom" which causes the quark bonds to become weak at high energies. Yang-Mills theory is shown to exhibit such behavior by Gross, Wilczek and Politzer, who received the Nobel Prize for their discovery in 2004.

The second important feature is color confinement of quarks and gluons. Confinement causes the force between color charged particles to increase with distance. We will discuss this behavior in detail later, in \autoref{sec:wilsonint}.

The third feature is having a ``mass gap". Mass gap is the difference between the vacuum energy and the lowest energy state, which is simply the mass of the lightest particle in the theory. Gluons are massless but strong interactions force them to form massive composite particles called glueballs, leading to the existence of a mass gap in the theory. The lightest force carrier being massive, clearly limits the range of the theory. If the mass gap has the value $\Delta$, then the theory is trivial at very large distances compared to $1/\Delta$.

In 3+1 dimensions, showing that the Yang-Mills action has these features turned out to be a very difficult task. This difficulty lead physicist to search for ways to simplify the problem.
\section{Why 2+1 Dimensions?}\label{sec:2+1}

Studies in 2+1 dimensional gauge theories mainly started in early 80s, after Feynman\cite{Feynman:1981ss} pointed out that QCD in 3+1 dimensions is too complicated to understand the qualitative behavior of the theory, since one needs to calculate 104 numbers at each point in space-time. To simplify the problem, he suggested a number of ways including decreasing the number of colors, decreasing the number of dimensions and studying just gluon systems without quarks. One or more of these methods can be used, depending on the qualitative feature one wants to understand. One of the most important of these qualitative features is confinement. Since even with no quarks there is a gluon confinement problem, studying just gluon systems is a nice and simpler way to study confinement. Also, switching from 3+1 D to 2+1 D significantly simplifies the problem. In 2+1 D with the gauge group SU(3), in a no quark system, one needs to calculate only 24 numbers at each point in space-time compared to 104 in the full problem. 1+1 D is exactly solvable but choosing it would be oversimplifying, because there are no propagating degrees of freedom. Although it is still mathematically interesting, this case would not be very helpful in studying the features of the theory that we are trying to understand.  But 2+1 D is non-trivial and has propagating degrees of freedom, which makes it more suitable for our interest.

An important progress in the 2+1 D program came from two groups. Karabali, Kim and Nair\cite{Karabali:1998yq} made an analytic prediction on the string tension(up to some approximation) and the other group, Teper, Lucini and Bringoltz\cite{teper1998n,lucini2002n,bringoltz2007precise} used lattice gauge theory Monte Carlo simulations to calculate it. Both groups had results that are in excellent agreement up to only 3\% difference. 

Three decades after Feynman's paper, a good understanding of 3+1 D Yang-Mills theory still seems out of reach. There have been important improvements in the 2+1 program but the next step is not clear. Our goal is to study the 2+1 D system with a Yang-Mills-Chern-Simons action to understand the transition of a topological theory to non-topological theory. Hopefully, this strategy will be helpful in filling a gap in the literature on 2+1 D gauge theories.
\section{Chern-Simons Theory}\label{sec:csint}

Chern-Simons(CS) theory is a topological field theory given by the action
\al{
S_{CS}=-\frac{k}{4\pi} \int {d^3x}\ \epsilon^{\mu\nu\alpha}\ Tr \pa{A_\mu \partial_\nu A_\alpha + \frac{2}{3}A_\mu A_\nu A_\alpha}.
}
This action was first studied in ref. \citen{chernsimons}. Chern and Simons observed that the Pontryagin density can be written as a total derivative,
\al{
\epsilon^{\mu\nu\alpha\beta}Tr (F_{\mu\nu}F_{\alpha\beta}) =4\ \partial_\beta\  \epsilon^{\mu\nu\alpha\beta}Tr \pa{A_\mu \partial_\nu A_\alpha + \frac{2}{3}A_\mu A_\nu A_\alpha}.
}
This boundary term was interesting in its own right and gave birth to the Chern-Simons Lagrangian.
Later, it was introduced to the physics literature by Jackiw and Templeton\cite{jackiw1981super}. 

Since the Lagrangian of this theory is metric free, the theory is topological. The equations of motion are given by
\al{
\label{eq:cseqmot}
\frac{\delta S_{CS}}{\delta A^a_\mu}=\frac{k}{8\pi}\epsilon^{\mu\nu\alpha}F^a_{\nu\alpha}=0.
}
Chern-Simon action is classically not gauge invariant. Under $A_\mu \rightarrow A^g_\mu=g A_\mu g^{-1} - \partial_\mu g g^{-1}$, the Lagrangian transforms as
\al{
\mc{L}_{CS} \rightarrow \mc{L}_{CS} + \frac{k}{4\pi} \epsilon^{\mu\nu\alpha} \partial_\mu\ Tr(\partial_\nu g g ^{-1} A_\alpha) + \frac{k}{12\pi} \epsilon^{\mu\nu\alpha}\ Tr(g^{-1}\partial_\mu g g^{-1}\partial_\nu g g^{-1}\partial_\alpha g).
}
The total derivative term can be made vanished by choosing suitable boundary conditions. The last term however, does not vanish. Up to a constant, the integral of this term is called the winding number $\omega(g)$, given by
\al{
\omega(g)=\frac{1}{24\pi^2}\int d^3x\ \epsilon^{\mu\nu\alpha}\ Tr(g^{-1}\partial_\mu g g^{-1}\partial_\nu g g^{-1}\partial_\alpha g).
}
With suitable boundary conditions, $\omega(g)$ is an integer. Now, we can write 
\al{
S_{CS}(A) \rightarrow S_{CS}(A^g) = S_{CS}(A) + 2\pi k \omega(g).
}
As we have stated before, Chern-Simons action is classically not gauge invariant but it can be made gauge invariant at the quantum level by constraining $k$ to be an integer. In that case, the weight of the path integral $e^{iS_{CS}}$ does not change, thus the theory becomes gauge invariant. The integer $k$ is usually referred to as the ``level number" of Chern-Simons theory. 

We will discuss geometric quantization of this theory in \eqref{ch:cs}.

\section{Topologically Massive Yang-Mills Theory}\label{sec:TMYMint}

Topologically massive Yang-Mills(TMYM) theory is a mixture of Chern-Simons(CS) and Yang-Mills(YM) theories.
The action is given by $S_{TMYM}=S_{CS}+S_{YM}$. As an introduction, we will only discuss the abelian case for now, following ref. \citen{nakahara}. The non-abelian theory will be discussed in \autoref{ch:tmym}. The abelian Lagrangian is given by 
\al{
\mc{L}_{TMYM}=-\frac{1}{4m}F_{\mu\nu}F^{\mu\nu}+\frac{1}{4}\epsilon^{\mu\nu\alpha}A_\mu F_{\nu\alpha}.
} 
Here $m$ is called the \emph{topological mass} for the following reason. The equations of motion are given by
\al{
\partial_\mu F^{\mu\nu}+\frac{m}{2}\epsilon^{\nu\alpha\beta}F_{\alpha\beta}=0
}
Using $\ast F^\mu=\frac{1}{2}\epsilon^{\mu\nu\alpha}F_{\nu\alpha}$ and $F^{\mu\nu}=\epsilon^{\mu\nu\alpha}\ast F_\alpha$ we can rewrite the equations of motion as
\al{
\epsilon^{\mu\nu\alpha}\partial_\mu \ast F_\alpha + m \ast F^\nu=0.
}
Multiplying both sides by the $\epsilon$ tensor gives,
\al{
\partial_\mu \ast F_\nu - \partial_\nu \ast F_\mu = m F_{\mu\nu}.
}
Then taking the covariant derivative of both sides and using the Bianchi identity $\partial_\mu \ast F^\mu=0$, we get
\al{
\label{eq:topmass}
(\partial_\mu \partial^\mu + m^2)\ast F_\nu=0.
}
It can be seen from \eqref{eq:topmass}, the additional topological(Chern-Simons) term in the Lagrangian gives mass to $\ast F_\nu$, hence the name topological mass.

\section{The Classical Equivalence}\label{sec:equiv}

As shown by Lemes et al.\cite{Lemes:1997vx},\cite{Lemes:1998md}, classically it is possible to write the TMYM action as a pure CS action,  via this nonlinear gauge field redefinition:
\al{
\label{eq:a-hat}
\hat{A}_\mu=A_\mu+\sum_{n=1}^{\infty}\frac{1}{m^n}\theta^n_\mu(D,F)
}
and the first three $\theta^n_\mu$ coefficients are
\vspace{0.20cm}
\al{
\label{eq:theta}
\bs
\theta^1_\mu=&\frac{1}{4}\epsilon_{\mu\sigma\tau}F^{\sigma\tau},\\
\theta^2_\mu=&\frac{1}{8}D^\sigma F_{\sigma\mu}~\text{and}\\
\theta^3_\mu=&-\frac{1}{16}\epsilon_{\mu\sigma\tau}D^\sigma D_\rho F^{\rho\tau}+\frac{1}{48}\epsilon_{\mu\sigma\tau}[F^{\sigma\rho},F^\tau_\rho].
\es
}
\vskip0.35cm\noindent
With this redefinition, the equivalence is given by
\al{
S_{CS}(\hat A)=S_{CS}(A)+S_{YM}(A)
}
where
\al{
S_{CS}(\hat A)=\frac{1}{2}\int_{\Sigma\times[t_i,t_f]} {d^3x }\ \epsilon^{\mu\nu\alpha}\ Tr \pa{\hat A_\mu \partial_\nu \hat A_\alpha + \frac{2}{3} \hat A_\mu \hat A_\nu \hat A_\alpha},
}
\al{
S_{CS}(A)=\frac{1}{2}\int_{\Sigma\times[t_i,t_f]} {d^3x }\ \epsilon^{\mu\nu\alpha}\ Tr \pa{A_\mu \partial_\nu A_\alpha + \frac{2}{3}A_\mu A_\nu A_\alpha}
}
and
\al{
S_{YM}(A)=\frac{1}{4m}\int_{\Sigma\times[t_i,t_f]} {d^3x }\ Tr\  F_{\mu\nu}F^{\mu\nu}.
}
This equivalence is classical and cannot be exactly extended to the quantum level. Some work on addressing this issue was done by Quadri\cite{Quadri:2002ni}, using the BRST formulation. 

Even though this equivalence does not work at the quantum level, it provides a good motivation to study the relation between TMYM theory and link invariants. 

\section{Topologically Massive AdS Gravity}\label{sec:TMG}

In three dimensions, there is a gravitational analogue of TMYM theory, called the topologically massive gravity model. In this section, following refs. \citen{witten1988,Carlip2008272, Carlip2}, we will study this gravity model with a negative cosmological constant, which is called the topologically massive AdS gravity model. For a dynamical metric $\gamma_{\mu\nu}$, this model has the action
\al{
\label{eq:tmgS}
S=\int d^3x \left[ -\sqrt{-\gamma}(R-2\Lambda)+\frac{1}{2\mu}\epsilon^{\mu\nu\rho}\left(  \Gamma^\alpha_{\mu\beta}\partial_\nu \Gamma^\beta_{\rho\alpha} +\frac{2}{3}\Gamma^\alpha_{\mu\gamma} \Gamma^\gamma_{\nu\beta} \Gamma^\beta_{\rho\alpha}   \right)  \right].
}
This action can be written as two CS terms with defining
\al{
A^{\pm}{}_{\mu}{}^a{}_b[e]=\omega_{\mu}{}^a{}_b[e] \pm \epsilon^a{}_{bc} e_\mu{}^c,
}
where $e_\mu{}^a$ is the dreibein and $\omega_{\mu}{}^a{}_b[e]$ is the torsion-free spin connection. Then, the action \eqref{eq:tmgS} can be written as
\al{
\label{eq:tmg2CS}
S[e]=-\frac{1}{2}\pa{1-\frac{1}{\mu}}S_{CS}\big[A^+[e]\big]+\frac{1}{2}\pa{1+\frac{1}{\mu}}S_{CS}\big[A^-[e]\big]
}
where 
\al{
S_{CS}[A]=\frac{1}{2}\int \epsilon^{\mu\nu\rho}\pa{A_\mu{}^a{}_b \partial_\nu A_\rho{}^b{}_a + \frac{2}{3} A_\mu{}^a{}_c A_\nu{}^c{}_b A_\rho{}^b{}_a}.
}
For our interests, the main difference between this gravity model and TMYM theory is that the latter has a mass gap, therefore it has a topological behavior only at large distances. But the gravity model is topological irrespective of the value of $\mu$. In TMYM theory, large distances are obtained by taking large values of $m$, which corresponds to small values of $\mu$ in the gravitational analogue. On the other hand, the $\mu\rightarrow\infty$ limit of the gravity model is analogous to pure YM theory. 

Now, let us focus on two important limits of \eqref{eq:tmg2CS}. For small values of $\mu$, \eqref{eq:tmg2CS} can be written as a \emph{sum} of two half CS theories as
\al{
\label{eq:tmgcs+cssplitting}
S[e]\approx\frac{1}{2\mu}S_{CS}\big[A^+[e]\big]+\frac{1}{2\mu}S_{CS}\big[A^-[e]\big].
}
In the $\mu\rightarrow\infty$ limit, it is equal to the \emph{difference} between two half CS theories, as
\al{
\label{eq:YMlimit}
S[e]=\frac{1}{2}S_{CS}\big[A^-[e]\big]-\frac{1}{2}S_{CS}\big[A^+[e]\big].
}
At large distances, it would be interesting to find out whether or not a similar CS+CS type splitting appears in TMYM theory and a CS--CS type splitting in pure YM theory. These cases will be discussed in \autoref{ch:tmym} and \autoref{ch:ym}. But we will not study pure YM theory as the small $m$ limit of TMYM theory, since this would force us to focus on short distances instead of large, where the scale dependence is strong and no topological behavior can be expected. Thus, we will study pure YM as a separate theory in \autoref{ch:ym}.
\section{The Wess-Zumino-Witten Model}\label{sec:WZW}

Wess-Zumino-Witten(WZW)\footnote{This section is a review, following ref. \citen{Nair:2005iw}.} model defines a conformal field theory in two dimensions. The action consists of two terms: a two dimensional kinetic term and a three dimensional topological term, as
\al{
S_{WZW}[U]=\frac{1}{8\pi}\int_{\partial \mc{M}}d^2x\ \sqrt g\ g^{ab} Tr(\partial_aU\partial_bU^{-1})+\Gamma[U],
}
where 
\al{
\Gamma[U]=-\frac{i}{12\pi}\int_{\mc{M}} d^3x\ \epsilon^{\mu\nu\alpha} Tr(U^{-1}\partial_\mu U U^{-1}\partial_\nu U U^{-1}\partial_\alpha U)
}
and $\mc{M}$ is a three dimensional space with a closed boundary $\partial\mc{M}$. $U(x)$ are invertible matrices that are not necessarily unitary. $\Gamma[U]$ is called the Wess-Zumino term. This term was added to the kinetic term by Witten, to bosonize non-abelian fields in two-dimensions\cite{Witten:1984uq}. The Wess-Zumino term is metric free, hence topological and its value is always an integer. 

In an Euclidean space, using complex coordinates $z=x_1-ix_2$ and $\bar{z}=x_1+ix_2$, we can rewrite the action as
\al{
\label{eq:wzwc}
S_{WZW}[U]=\frac{1}{2\pi}\int_{\partial \mc{M}} Tr(\partial_z U \partial_{\bar{z}}U^{-1})+\Gamma[U].
}
The equations of motion for this model are given by
\al{
\partial_z(U^{-1}\partial_{\bar{z}}U)=0\ ~\text{and}~\ \partial_{\bar{z}}(\partial_z UU^{-1})=0,
}
which defines two currents $J_z=-\frac{1}{\pi}\partial_z UU^{-1}$ and $J_z=\frac{1}{\pi}U^{-1}\partial_{\bar{z}}U$. These currents are associated with infinitesimal left and right transformations $U\rightarrow (1+f(z))U$ and $U\rightarrow U(1+g(\bar{z}))$. The action \eqref{eq:wzwc} is invariant under these transformations.

The Wess-Zumino term has the following useful property,
\al{
\label{eq:gamma}
\Gamma[UM]=\Gamma[U]+\Gamma[M]-\frac{i}{4\pi}\int_{\partial\mc{M}} d^2x\ \epsilon^{ab}Tr(U^{-1}\partial_a U M^{-1}\partial_b M).
}
Equation \eqref{eq:gamma} leads to a very important identity called the Polyakov-Wiegmann(PW)\cite{Polyakov1983121,Polyakov1984223} identity, given by
\al{
S_{WZW}[UM]=S_{WZW}[U]+S_{WZW}[M]+\frac{1}{\pi}\int_{\partial\mc{M}} Tr(U^{-1}\partial_zU\partial_{\bar{z}}MM^{-1}).
}
A chiral splitting can be seen here, since there is only the anti-holomorphic derivative of $U$ and only the holomorphic derivative of $M$.

The WZW action can be used to calculate two-dimensional fermion determinants. Massless fermion Lagrangian naturally splits into two chiral terms as,
\al{
\mc{L}=\bar{\psi}D_z\psi+\bar{\chi}D_{\bar{z}}\chi
}
where $\psi$ and $\chi$ are chiral components of a spinor and $D_i$ is the covariant derivative in the adjoint representation, given by $D_i=\partial_i+A_i$. $det D_z$ and $det D_{\bar{z}}$ can be evaluated using the WZW action. We will start by parametrizing the components of a gauge field $A$ with a complex matrix $U$. A good starting point would be taking $U$ to be something similar to
\al{
\label{eq:U0}
U(x,0,C)=\mc{P}\ exp\pa{-\underset{C}{\ \ \int_0^x} A_i dx^i},
}
where $\mc{P}$ is the path ordering operator. Then the gauge field can be written as $A_i=-\partial_i U U^{-1}$. However, this is not a good way to parametrize the gauge fields, since it depends on the path $C$. A small deformation at point $x_0$ on the path results in a change
\al{
\delta U(x,0,C)=\mc{P}\ exp\pa{-\underset{C}{\ \ \int_{x_0}^x} A_i dx^i}\ F_{ij}\Sigma^{ij}\ \mc{P}\ exp\pa{-\underset{C}{\ \ \int_0^{x_0}} A_i dx^i},
}
where $F_{ij}=\partial_i A_j - \partial_j A_i +[A_i,A_j]$ and $\Sigma^{ij}$ is the area element. To make the matrix path independent, $F_{ij}$ must be made vanished. In Chern-Simons theory this condition is satisfied on shell but Yang-Mills theory has no such behavior. To make sure that the curvature vanishes irrespective of the Lagrangian, $U$ can be chosen as
\al{
\label{eq:U}
U(x,0,C)=\mc{P}\ exp\pa{-\underset{C}{\ \ \int_0^x} \mc{A}_z dz+A_{\bar{z}}d\bar{z}},
}
where $\mc{A}_z$ is defined by $\partial_z A_{\bar{z}} - \partial_{\bar{z}} \mc{A}_z +[\mc{A}_z,A_{\bar{z}}]=0$. This flatness condition forces $\delta U$ to be zero under small deformations of the path $C$. With this definition, the gauge fields can be parametrized as
\al{
\label{eq:para}
A_{\bar{z}}=-\partial_{\bar{z}}UU^{-1}\ ~\text{and}~\ A_z=U^{\dagger-1}\partial_z U^\dagger.
}
This parametrization is sometimes referred to as ``Karabali-Nair parametrization"\cite{Karabali1996135} in the literature. Also, $\mc{A}_i$ can be written as
\al{
\mc{A}_{\bar{z}}=U^{\dagger-1}\partial_{\bar{z}} U^\dagger\ ~\text{and}~\ \mc{A}_z=-\partial_z UU^{-1}.
}  
The flatness condition $\partial_z A_{\bar{z}} - \partial_{\bar{z}} \mc{A}_z +[\mc{A}_z,A_{\bar{z}}]=0$ can be rewritten as $\mc{A}_z=(D_{\bar{z}}^{-1})\partial_z A_{\bar{z}}$, which implies that the covariant derivative must be invertible. This requirement introduces a restriction on the topology of the space. In simply connected spaces, $D_{\bar{z}}$ does not have zero modes. However, when the space is not simply connected, for example $\Sigma=S^1\times S^1$,  parametrization \eqref{eq:para} does not work. In this case the correct parametrization is
\al{
A_{\bar{z}}=-\partial_{\bar{z}}UU^{-1}+U\pa{\frac{i\pi a}{Im\tau}}U^{-1},
}
where $a$ is a constant gauge field and $\tau$ is the modular parameter of the torus.

Matrices $U$ and $U^\dagger$ transform like fermions, as
\al{
\label{eq:Utransform}
U\ \rightarrow\ gU \ ~\text{and}~\ U^\dagger \rightarrow\ U^\dagger g^{-1}.
}

To write 2D fermion determinants in terms of the WZW action, let us define an action as $\mc{S} \equiv log\ det\ D_{\bar{z}}=Tr\ log D_{\bar{z}}$. After doing point-splitting regularization on $D_{\bar{z}}^{-1}$ \cite{Nair:2005iw}, the variation of the action can be written as
\al{
\label{eq:deltaS}
\delta \mc{S}=-\frac{1}{\pi}\int d^2x\ Tr(\mc{A}_z\delta A_{\bar{z}})_A,
}
where label $A$ indicates the adjoint representation. At this point we need to switch to the fundamental representation using $Tr(t^at^b)_A=2c_A\ Tr(t^at^b)_F$, where $c_A$ is the quadratic Casimir in the adjoint representation given by $c_A \delta^{ad}=f^{abc}f^{dbc}$. This allows us to write \eqref{eq:deltaS} as the variation of $S_{WZW}(U)$ up to a constant, as
\vspace{0.20cm}
\al{
\bs
\delta \mc{S}=&-\frac{2c_A}{\pi}\int d^2x\ Tr(\mc{A}_z\delta A_{\bar{z}})_F\\
=&2c_A\ \delta S_{WZW}(U).
\es
}
\vskip0.35cm\noindent
Now we can write $det D_{\bar{z}}$ in terms of $S_{WZW}$ as
\al{
\label{eq:det1}
det D_{\bar{z}}=det (\partial_{\bar{z}})\ exp\big(2c_A S_{WZW}(U)\big)
}
and $det D_z$ is the complex conjugate of $det D_{\bar{z}}$, written as
\al{
\label{eq:det2}
det D_z=det (\partial_z)\ exp\big(2c_A S_{WZW}(U^\dagger)\big).
}
These determinants are not gauge invariant. Under infinitesimal gauge transformations, the action $\mc{S}$  transforms as
\al{
\label{eq:wzwtransform}
\delta_\epsilon \mc{S}=-\frac{1}{\pi}\int d^2x\ Tr(\partial_z A_{\bar{z}}\ \delta g  g^{-1}) = \frac{1}{2\pi}\int d^2x\ \varepsilon^a\ \partial_z A^a_{\bar{z}}
}
where $A_\mu=-iA_\mu^a t^a$, $Tr(t^a,t^b)_F=\frac{1}{2}\delta^{ab}$ and $ g\approx 1-it^a\varepsilon^a$ is used $(\varepsilon<<1)$.  

Using \eqref{eq:det1} and \eqref{eq:det2}, the full Dirac determinant can be written as
\al{
det (D_{\bar{z}}D_z)=det (\partial_{\bar{z}}\partial_z)\ exp\big[2c_A \big(S_{WZW}(U)+S_{WZW}(U^\dagger)\big)\big],
}
which is still not gauge invariant. A local counter term can be added to the exponential to complete the PW identity. Adding a local counter term is allowed, since it is equivalent to choosing a different regularization. The complete PW identity is
\al{
S_{WZW}(U^\dagger U)=S_{WZW}(U)+S_{WZW}(U^\dagger)+\frac{1}{\pi}\int d^2x\ Tr(\partial_{\bar{z}}UU^{-1}U^{\dagger-1}\partial_z U^\dagger).
}
The last term is the local counter term that we have added, which can be written as $-\frac{1}{\pi}\int d^2x\  Tr(A_{\bar{z}}A_z )$. Now we have everything we need to write the full Dirac determinant as a gauge invariant function. Since $det (\partial_{\bar{z}}\partial_z)$ is just a constant factor\cite{Bos:1989kn}, the full Dirac determinant can be written as
\al{
\label{eq:diracdet}
det (D_{\bar{z}}D_z)=constant \times exp\big(2c_A S_{WZW}(H)\big),
}
where $H=U^\dagger U$. It can be seen from \eqref{eq:Utransform} that $H$ is $SU(N)$ gauge invariant, so is the Dirac determinant \eqref{eq:diracdet}.

\section{Wilson Loops}\label{sec:wilsonint}

\begin{figure}[H]
\centering
\includegraphics[scale=0.7]{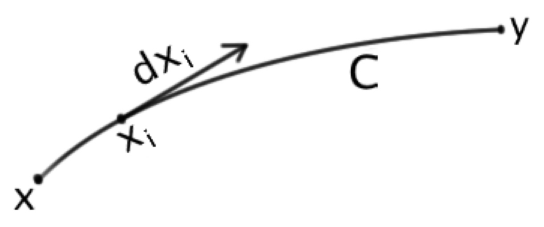}
\caption{Path $C$}
\label{fg:path}
\end{figure}

One of the most important observables in gauge theory is the gauge invariant Wilson Loop, which is given by the holonomy of the gauge field around a space-time loop. This observable is non-local, like electromagnetic flux. Following refs. \citen{Nair:2005iw, ramond}, construction of this observable can be done as follows. We look at the parallel transport of matter fields. Let $\phi$ be a matter field. Then, analogous to Riemannian geometry, its parallel transport can be analyzed with the covariant derivative as
\al{
D_\mu\phi=(\partial_\mu+A_\mu)\phi=0.
}
This equation can be solved by $\phi=U\phi_0$ where $\phi_0$ is a constant and $U$ is given by
\al{
\label{eq:cmu}
C^\mu(\partial_\mu+A_\mu)U=0,
}
where $C^\mu$ are components of a vector tangent to a curve $C$. The reason for introducing a curve here is the non-existence of a general well-defined solution to the equation $(\partial_\mu+A_\mu)U=0$.
The equation \eqref{eq:cmu} is solved by
\vspace{0.20cm}
\al{
\bs
U(y,x,C,A_\mu)=&\underset{i}\prod \big(1- A_\mu(x_i)dx_i^\mu\big) \\
=&\mc{P}exp\pa{-\underset{C}{\ \ \int_x^y}A_\mu dx^\mu},
\es
}
\vskip0.35cm\noindent
where $\mc{P}$ is the path ordering operator and definitions of $x_i$ and $dx_i$ can be seen clearly in \autoref{fg:path}.
\vspace{-.20cm}
\noindent
$Tr(U(y,x,C,A_\mu))$ is the definition of the Wilson line operator and it is neither gauge invariant nor covariant. Under $A\rightarrow A^g=gAg^{-1}-dgg^{-1}$, $U(y,x,C,A_\mu)$ transforms like
\al{
U(y,x,C,A^g_\mu)=g(y)U(y,x,C,A_\mu)g(x)^{-1}.
}
To build a gauge invariant quantity, instead of integrating over an open path, we can integrate over a closed loop, where $x$ and $y$ are the same point. Trace of $U(x,x,C,A_\mu)$ is gauge invariant and can be written as
\al{
W_R(C)=Tr_R\ \mc{P}\ exp\pa{-\underset{C}\oint A_\mu dx^\mu} .
}
$W_R(C)$ is called the Wilson loop operator and subscript $R$ indicates the representation dependence of this observable.

\subsection{Confinement}

Wilson loop is an important tool to study confinement. If the theory is confined, the Wilson loop expectation value should satisfy the \emph{area law}
\al{
\label{eq:area}
\langle W_R(C)\rangle =e^{-\sigma A_C},
}
where $\sigma$ is a constant and $A_C$ is the area of the loop, given by the closed curve $C$. For a non-confining theory, the expectation value may follow the \emph{perimeter law}
\al{
\label{eq:perimeter}
\langle W_R(C)\rangle =e^{-mL_C},
}
where $m$ is a constant and $L_C$ is the circumference of the loop. For a theory to have either \eqref{eq:area} or \eqref{eq:perimeter} type of behavior, there has to be a mass gap since both laws follow a decay pattern at large scales.

To see how Wilson loop is related to confinement, following ref. \citen{greiner2007quantum}, we will take a look at a quark-antiquark pair. The potential between a quark and an antiquark depends linearly on the separation $R$, except for very small values of $R$. For this case, the potential can be written as
\al{
\label{eq:V}
V_{q\bar{q}}(R)\overset{R\rightarrow\infty}\longrightarrow\sigma R,
}
$\sigma$ is a constant called the \emph{string tension}. This behavior is explained by assuming a flux tube connecting the two quarks. The color analogue of electric field lines are confined in a tube-like region which acts like a string, connecting the quark-antiquark pair. 

Confinement can be visualized with the help of an analogous electromagnetic system. Let us consider a hypothetical magnetic monopole-antimonopole pair that is placed inside an electric superconductor. 
\begin{figure}[h!]
\centering
\includegraphics[scale=0.47]{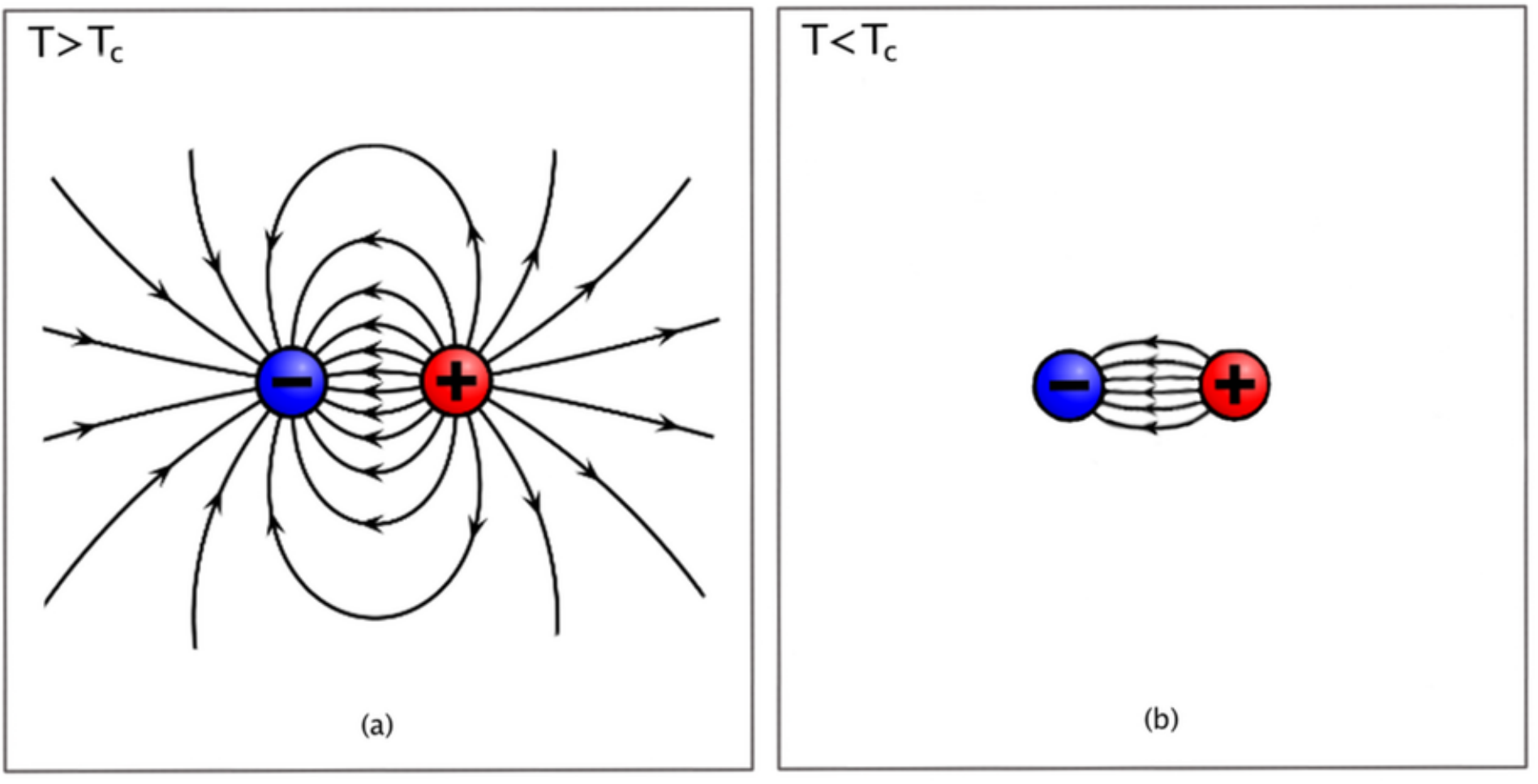}
\captionsetup{width=15cm}
\caption{A magnetic monopole-antimonopole pair placed in an electric superconductor.}
\label{fg:confine}
\end{figure}\vspace{-.20cm}
\noindent 
In \autoref{fg:confine} (a), the conductor is not in the super-conducting state, thus the magnetic field lines form the usual dipole pattern. \autoref{fg:confine} (b) shows what happens to the field lines in the superconducting state. As a result of the Meissner effect, the magnetic field lines are compressed, forming a flux tube. What happens in QCD is analogous to this example, with interchanging the magnetic field lines with electric field lines and replacing the electric superconductor with a magnetic analogue. This explanation of confinement is known as the ``dual super-conductor model"\cite{ripka2004dual}.  

\begin{figure}[H] 
\centering
\includegraphics[scale=0.40]{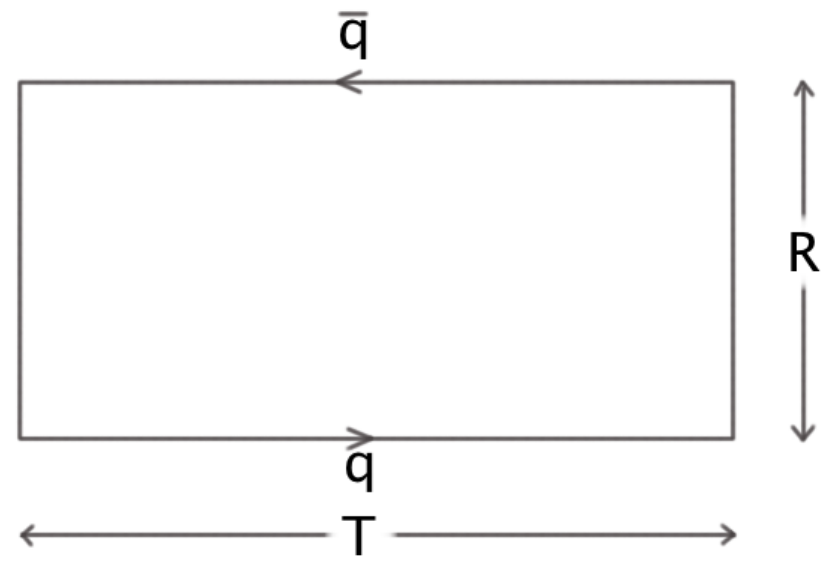}
\captionsetup{width=5.5cm}
\caption{A rectangular loop for a quark-antiquark pair.}
\label{fg:loop}
\end{figure}

Now let us assume that the quark-antiquark pair($q\bar{q}$) is heavy enough, so their kinetic energy can be neglected. In this case, the action is
\al{
S=-\int_{0}^{T} dt\ V_{q\bar{q}}(R)=TV_{q\bar{q}}(R).
}
This action is given by a quark-gluon interaction, thus it can be written as $S=-\int d^4x\ A_\mu j^\mu$. Since we assumed that the quark-antiquark pair is almost static, we can use the color charge density $\rho(x)=\delta^3(x)-\delta^3(x-R)$ and write
\al{
j^\mu(x) A_\mu(x)=\rho(x) A_0(x)= \pa{\delta^3(x)-\delta^3(x-R)} A_0(x).
}
Now, the action can be written as
\al{
S=-\int d^4x\  A_\mu j^\mu=-\int_0^T dt\big(A_0(0,t)-A_0(R,t)\big)
}
which should be equal to $T V_{q\bar{q}}(R)$, if the theory is confined. To relate this to a Wilson loop, we take the rectangular loop shown in \autoref{fg:loop}. 
\vspace{-.20cm}
\noindent
If the measurement is taken for a long time, then $T \gg R$ can be used for approximation. With this approximation, the loop integral can be written as
\vspace{0.20cm}
\al{
\bs
-\oint A_\mu dx^\mu =& - \int_0^T dt A_0(0,t) +\int_0^R dx A_x(x,T)\\
 &+ \int_T^0 dt A_0(R,t) + \int_R^0 dx A(x,0)\\
\overset{T \gg R}\approx & T V(R).
\es
}
\vskip0.35cm\noindent
With $V_{q\bar{q}}(R)=\sigma R$, the Wilson expectation value is given by
\al{
\langle W(C) \rangle = \left\langle{ Tr\ \mc{P}\ exp\pa{-\underset{C}\oint A_\mu dx^\mu} }\right\rangle \approx e^{-\sigma RT}.
}
Since $RT$ is the area of the loop, we obtain the area law $\langle W_R(C)\rangle \approx e^{-\sigma A_C}$ for $T \gg R$.

In some non-confining theories, Wilson loop expectation values may follow the perimeter law. This is not the case for topological theories, since they do not have a mass gap. They cannot satisfy a decay law due to their scale independence. For example, Chern-Simons theory Wilson loop expectation values are given by link invariant polynomials that depend only on the topology of a knot or link. Thus, the area or circumference of the loop is irrelevant. This will the subject of the \autoref{sec:knot}. 

\subsection{'t Hooft Loop Operator} 

't Hooft loop operator is a dual version of the Wilson loop operator. It is defined in a similar way to Wilson loop with replacing the gauge field $A_\mu$ with its dual $\tilde{A}_\mu$, which is given by $E^i=\epsilon^{ijk}\partial_j \tilde{A}_k$. The 't Hooft loop operator is given by
\al{
\label{eq:thooftloop}
T(C)=Tr\ \mc{P}\ exp \pa{ -\underset{C}\oint \tilde A_\mu dx^\mu}.
}
This operator, together with the Wilson loop in fundamental representation, satisfies the 't Hooft algebra, 
\al{
\label{eq:thooftalgebra}
W_F(C)T(C')=e^{\frac{2\pi i}{N}L(C,C')}\ T(C')W_F(C).
}

\section{Knot Theory}\label{sec:knot}

The mathematical definition of a knot is, a smooth embedding of a circle in a 3 or higher dimensional space. A finite union of non-intersecting knots is called a link. If a knot can be continuously deformed into another without crossing itself, these two knots are said to be equivalent. 

\subsection{History of Knot Theory}

The history of knot theory dates back to 1867, when Lord Kelvin introduced his idea that atoms are knotted vortex tubes of ether\cite{atiyah1990geometry}. P.G. Tait, a collaborator of Kelvin's, was the first to study the classification of knots and his discoveries has been called ``Tait conjectures". An important improvement in knot theory was the discovery of the first link invariant polynomial in 1928: the ``Alexander Polynomial"\cite{alexander1928topological}. This invariant was not suitable enough to prove most of Tait conjectures. A more useful link invariant ``Jones polynomial" was discovered in 1984 by Vaughan Jones\cite{jones1985polynomial}, which made proving some of Tait conjectures possible\cite{atiyah1990geometry}. 

Knot theory was disregarded by the physics community for a long time after Kelvin's atom theory was proven wrong. But in late 1980s, Witten's work reintroduced knot theory to physics. Starting with Michael Atiyah's proposal\cite{atiyah}, Witten showed that Chern-Simons theory Wilson loop expectation values can be described by the Jones polynomial or related invariants. This discovery opened a new path in understanding gauge theories and conformal field theories.

\subsection{The Jones Polynomial and Skein Relations}\label{sec:jones}

The Jones polynomial of a link is a polynomial of a parameter $t$, denoted as $V_L(t)$. Jones polynomial of the mirror image of a knot is given by the transformation $t\rightarrow t^{-1}$; this is not the case for the Alexander polynomial since it cannot distinguish the knot from its mirror image. 

\begin{figure}[H]
\centering
\includegraphics[scale=0.5]{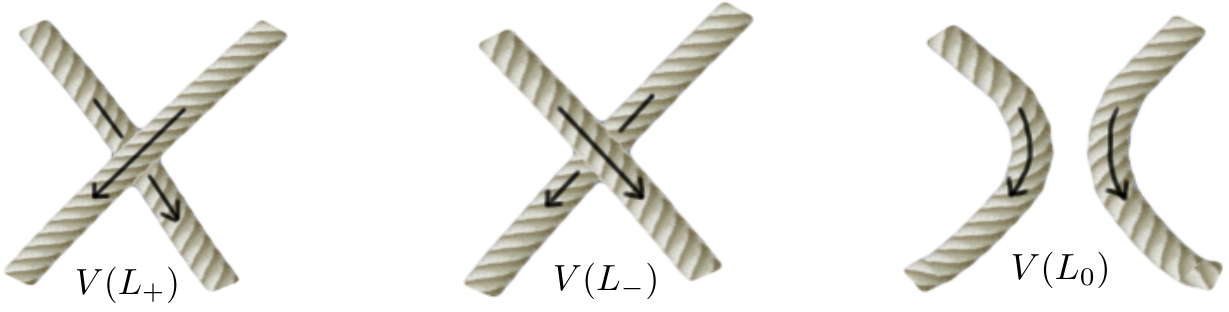}
\captionsetup{width=10.5cm}
\caption{Skein Relations: Three different versions of a crossing on a link is depicted above. $V_{L_+}, V_{L_-}$ and $V_{L_0}$ are the polynomials of the link with the corresponding version of the crossing.}
\label{fg:skein}
\end{figure}

The main tool to calculate the Jones polynomial is called a skein relation. To use a skein relation, one needs to take the projection of a three-dimensional link on a two-dimensional surface. This leads to having crossings on the projection. Using a skein relation is basically doing surgery on one of these crossings. Jones polynomials satisfy the following skein relation,
\al{
\label{eq:jonesskein}
t^{-1}V_{L_+}+tV_{L_-}=(t^{1/2}-t^{-1/2})V_{L_0}
}
with the knot diagrams given in \autoref{fg:skein}. 
\vspace{-.20cm}

The skein relation can be used recursively to calculate the polynomials of complicated links, starting with the normalization condition for the unknot(or simply, a circle) $V_o(t)=1$.

The Jones polynomial of a link that consists of an odd number of knots is always a Laurent polynomial of $\sqrt{t}$; for an even number, it is a polynomial of $t$.

\subsection{An Example: Jones Polynomial of a Trefoil Knot}

\begin{figure}[H]
\centering
\includegraphics[scale=0.6]{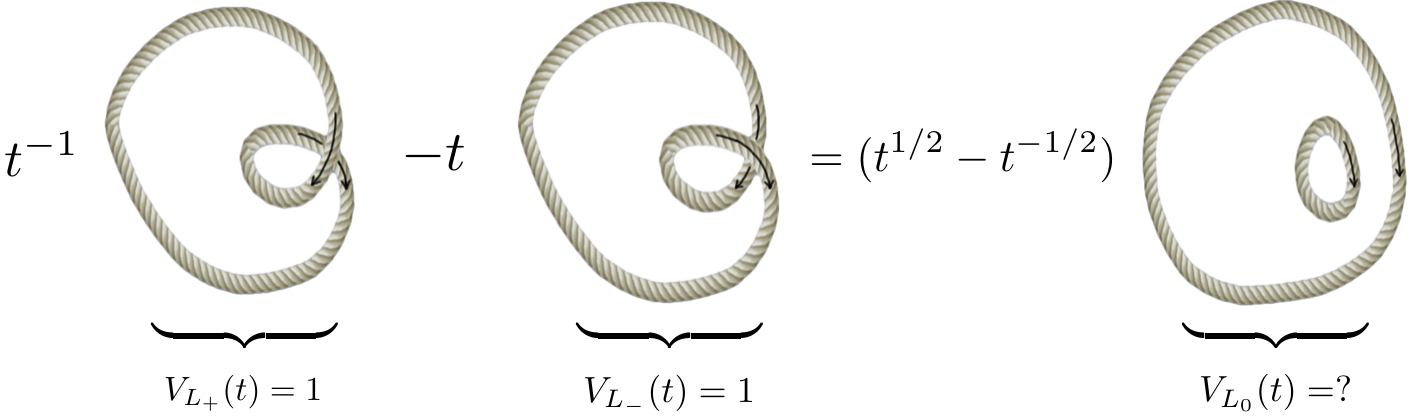}
\captionsetup{width=14.5cm}
\caption{A skein relation to find the Jones polynomial of a link that consists of two separate unknots.}
\label{fg:knots1}
\end{figure}

\begin{figure}[H]
\centering
\includegraphics[scale=0.6]{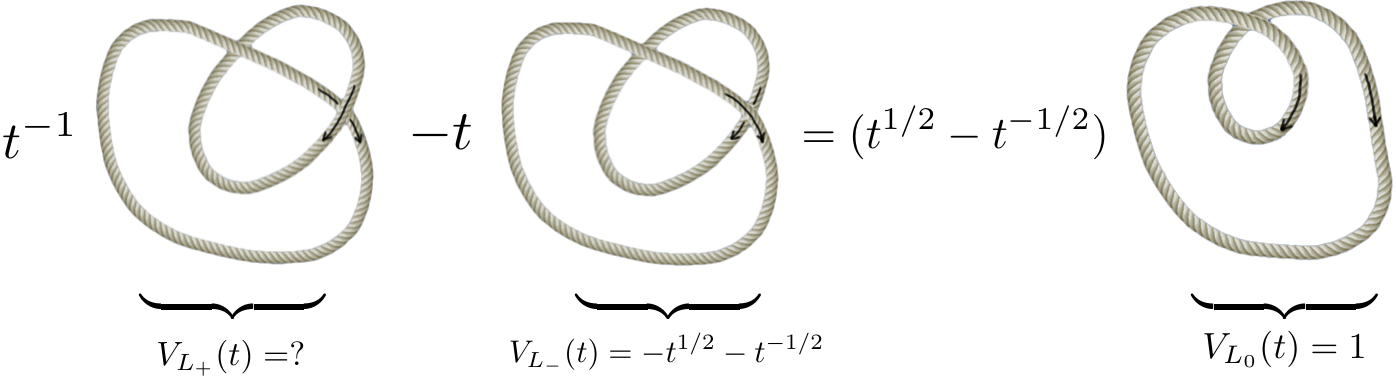}
\caption{A skein relation to find the Jones polynomial of a Hopf link.}
\label{fg:knots2}
\end{figure}

Now as an example, we will calculate the Jones polynomial for the right-handed trefoil knot. We start with two twisted unknots to get the polynomial of a link that consists of two separate unknots, as shown in \autoref{fg:knots1}. In all upcoming figures of this section, the  projection of the link represents the polynomial. Surgery will be done only on the crossing with arrows on it, which indicate the orientation of the knots. An orientation is necessary, since skein relations are meaningless without one. Without an orientation, the first two diagrams in \autoref{fg:skein} would be topologically equivalent, since one of them is just the rotated version of the other one.
\vspace{-.20cm}
\noindent 
Solving the skein relation in \autoref{fg:knots1} for $V_{L_0}$, we get
\al{
V_{L_0}(t)=-t^{1/2}-t^{-1/2}.
}
The next polynomial we need, is for the Hopf link, which consists of two unknots that go through each other as shown in the first term of \autoref{fg:knots2}.
\vspace{-.20cm}
\noindent
Notice that in \autoref{fg:knots2}, the link in the middle is the same two unknots we calculated in \autoref{fg:knots1}, and the last one is just a twisted unknot. Solving for $V_{L_+}$ gives,
\al{
V_{L_+}(t)=-t^{5/2}-t^{1/2}.
}
\begin{figure}[H]
\centering
\includegraphics[scale=0.6]{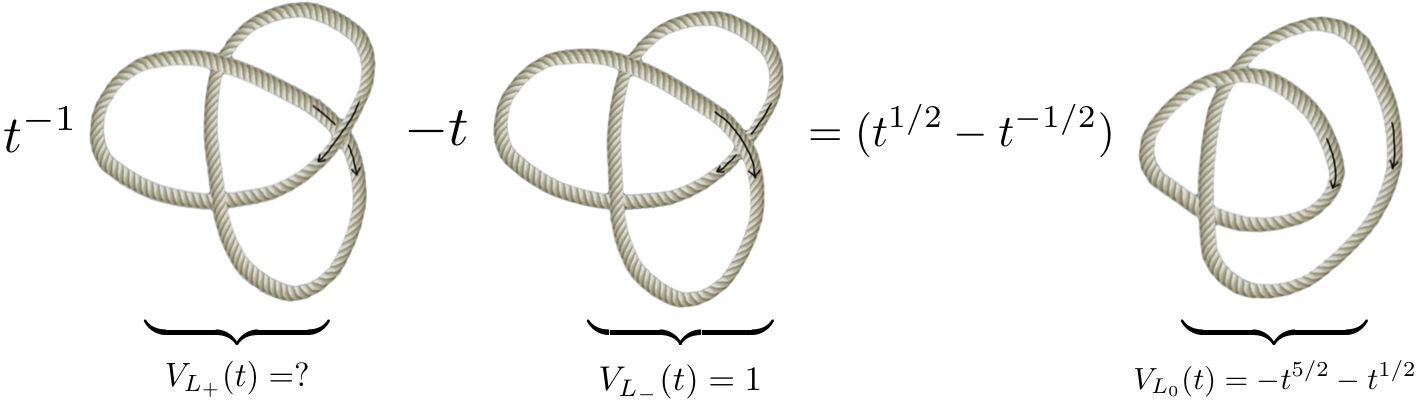}
\captionsetup{width=14.5cm}
\caption{A skein relation to find the Jones polynomial of a right-handed trefoil knot.}
\label{fg:knots3}
\end{figure}\vspace{-.20cm}
\noindent
Now we have everything we need to calculate the Jones polynomial for the trefoil knot, which is the first term in \autoref{fg:knots3}. The middle knot in \autoref{fg:knots3} is an unknot and the last link is the Hopf link that we calculated with the previous skein relation. Solving this skein relation for $V_{L_+}$ we get,
\al{
V_{trefoil}(t)=t+t^3-t^4.
}

\subsection{The HOMFLY Polynomial}

\begin{figure}[h!]
\centering
\includegraphics[scale=0.6]{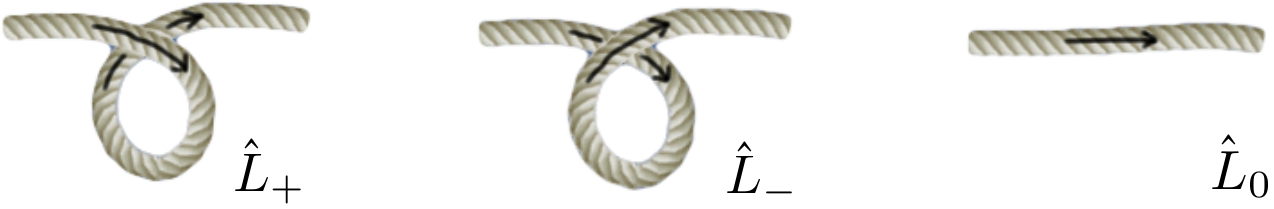}
\caption{Knot diagrams for $\hat L$.}
\label{fg:skein2}
\end{figure}

There is a generalized version of the Jones polynomial, called the ``HOMFLY" polynomial. This polynomial was discovered by Jim Hoste, Adrian Ocneanu, Kenneth Millett, Peter J. Freyd, W. B. R. Lickorish, and David N. Yetter\cite{freyd1985new}. The HOMFLY polynomial $S_L(\beta, z)$ satisfies the ``generalized" skein relations
\vspace{0.20cm}
\al{
\label{eq:genskein}
\bs
S_{\hat L_+} &= \alpha\ S_{\hat L_0},\\
S_{\hat L_-} &= \alpha^{-1} S_{\hat L_0},\\
\beta S_{L_+} - \beta &^{-1} S_{L_-} =z\ S_{L_0}.
\es
}
\vskip0.35cm\noindent
Here, $\beta$ is a parameter analogous to $t$ in \eqref{eq:jonesskein}. $z$ and $\alpha$ are functions of $\beta$. The normalization condition is $S_0=1$ for the unknot, similar to the Jones polynomial. However, the knot diagrams for $\hat L$ differ from \autoref{fg:skein}. The correct knot diagrams are given in  \autoref{fg:skein2}.
\vspace{-.20cm}
Although it looks like the diagrams for $\hat L_+$ and $\hat L_-$ are topologically the same in \autoref{fg:skein2}, they are not necessarily. Other parts of the knot can go though the loop, which is not shown in the figure. Solving \eqref{eq:genskein} is very similar to solving \eqref{eq:jonesskein} as shown in the example in \autoref{sec:jones}.

In the next section we will derive generalized skein relations for Wilson loop expectation values in Chern-Simons theory.

\section{Chern-Simons Theory and Link Invariants}

After Atiyah's proposal of the problem in 1988\cite{atiyah}, Witten found the meaning of the Jones polynomial in quantum field theory. Using 2D CFTs\cite{Witten:1988hf}, he showed that Chern-Simons theory Wilson loop expectation values are link invariants. But here, for pedagogical reasons, we will review Cotta-Ramusino et al.'s work\cite{CottaRamusino:1989rf} that used only 3D field theory techniques.

As we defined earlier, a link is a union of $n$ non-intersecting knots. In the physics language this is a product of Wilson loops 
\al{
\langle W_{R_1}(C_1) \dots W_{R_n}(C_n) \rangle \equiv \langle W(L) \rangle,
}
where each loop is a knot. 

To derive skein relations, we need to focus on crossings. $L_+$, $L_-$ and $L_0$ differ by small variations of the path on the same crossing. Under such a variation at a point $y$, the matrix $U(x_2,x_1)=\mc{P}\ exp \big(-\underset{C}{\int_{x_1}^{x_2}} A_\mu dx^\mu \big)$ changes by
\al{
\label{eq:deltU}
\delta U(x_2,x_1,C)=iU(x_2,y,C)\ F^a_{\mu\nu}\Sigma^{\mu\nu}t_R^a\ U(y,x_1,C),
}
where $\Sigma_{\mu\nu}$ is the area element of the deformed region, label $R$ indicates representation and there is no sum over $\mu$ and $\nu$. Since we want to see how $\langle W(L) \rangle$ changes under a variation of the path, we would like to calculate the expectation value $\langle F^a_{\mu\nu}(y) O_1 O_2 \dots \rangle $. Using \eqref{eq:cseqmot}, we can replace $F_{\mu\nu}$ by
\al{
\label{eq:F-CSeqmot}
F^a_{\mu\nu}=\frac{4\pi}{k}\epsilon_{\alpha\mu\nu} \frac{\delta S_{CS}}{\delta A^a_\alpha}.
}
With $Z = \int \mc{D}A\ e^{iS_{CS}}$, we write
\al{
\langle F^a_{\mu\nu}(y)O_1 O_2 \dots \rangle = \frac{1}{Z}\int \mc{D}A\ e^{iS_{CS}} F^a_{\mu\nu}(y)O_1 O_2 \dots.
}
Using \eqref{eq:F-CSeqmot}, the expectation value can be written as
\al{
\langle F^a_{\mu\nu}(y)O_1 O_2 \dots \rangle = -i\frac{4\pi}{k} \frac{1}{Z}\int \mc{D}A\ \epsilon_{\alpha\mu\nu} \frac{\delta e^{iS_{CS}}}{\delta A^a_\alpha}O_1 O_2 \dots.
}
After integrating by parts, the expectation value becomes
\al{
\label{eq:FOO}
\langle F^a_{\mu\nu}(y)O_1 O_2 \dots  \rangle = i\frac{4\pi}{k} \frac{1}{Z}\int \mc{D}A\ e^{iS_{CS}}\epsilon_{\alpha\mu\nu} \frac{\delta }{\delta A^a_\alpha}(O_1 O_2 \dots ).
}
Now, using \eqref{eq:deltU} and \eqref{eq:FOO}, we can calculate how $\langle \dots U(x_2,x_1) \dots \rangle$ changes under a small variation of the path. The result is
\al{
\delta \langle \dots U(x_2,x_1) \dots \rangle = -i\frac{4\pi}{k}\delta^3(x-y)\epsilon_{\alpha\mu\nu}\Sigma^{\mu\nu}dy^\alpha \langle \dots U(x_2,x)  t_R^a t_R^a  U(y,x_1) \dots \rangle.
}
The factor in front of the term on the right hand side plays the key role in obtaining skein relations. Without the constants, the factor is $v=\delta^3(x-y)\epsilon_{\alpha\mu\nu}\Sigma^{\mu\nu}dy^\alpha$ and it can only take values $0, \pm 1$. $dy^\alpha$ is tangent to the path and its orientation with respect to the plane defined by $\Sigma^{\mu\nu}$ decides the value of $v$. If $dy^\alpha$ is on the $\Sigma^{\mu\nu}$ plane, then the Levi-Civita tensor will take the value zero, since two of the indices will be the same. When $dy^\alpha$ is perpendicular to the plane, then depending on the direction $v$ takes values $\pm1$. This leads to three possible deformations of the a link $L$, as
\al{
\label{eq:deltW1}
\delta \langle W_R(L) \rangle=0
}
and
\al{
\label{eq:deltW2}
\delta \langle W_R(L) \rangle= \pm i \frac{4\pi}{k} c_R\langle W_R(L) \rangle
}
where $c_R \mathds{1}=\underset{a}\sum t_R^a t_R^a $ is the quadratic Casimir in representation $R$. To cross over or under, the deformation has to be perpendicular to the plane defined by $\Sigma^{\mu\nu}$. Thus, in \autoref{fg:skein2}, $L_+$ and $L_-$ corresponds to $dy^\alpha$ being perpendicular to the plane, which is given by \eqref{eq:deltW2}. The case where $dy^\alpha$ belongs to the plane is given by \eqref{eq:deltW1}. 

\begin{figure}[H]
\centering
\includegraphics[scale=0.7]{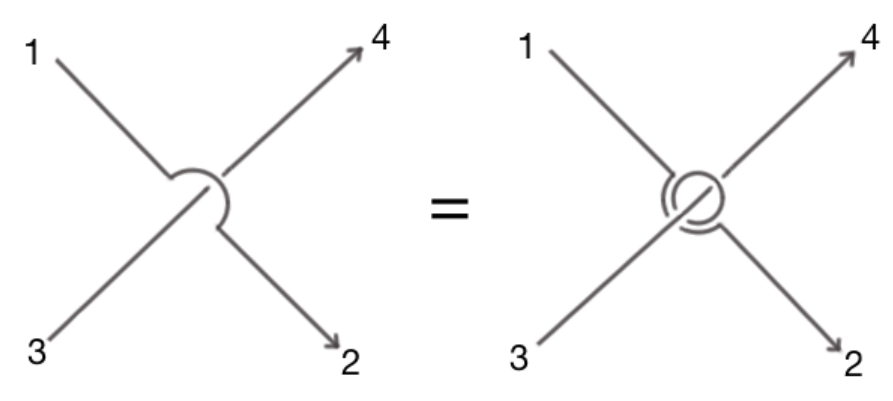}
\caption{Relating under-crossings with over-crossings.}
\label{fg:cross}
\end{figure}

The expectation values satisfy the following recursive relation,
\al{
\label{eq:skeinlike}
\langle W(\hat L_+)\rangle - \langle W(\hat L_-)\rangle =-i\frac{4\pi}{k}c_R \langle W(\hat L_0)\rangle.
}
Since our goal is to obtain \eqref{eq:genskein}, we write
\al{
\label{eq:Walpha}
\langle W(\hat L_+)\rangle = \alpha \langle W(\hat L_0)\rangle \ ~\text{and}~\ \langle W(\hat L_-)\rangle = \alpha^{-1} \langle W(\hat L_0)\rangle.
}
Solving \eqref{eq:skeinlike} using \eqref{eq:Walpha} gives
\al{
\alpha = 1 - i\frac{2\pi}{k}c_R +\mc{O}\pa{\frac{1}{k^2}}.
}
Deforming with an infinitesimally small circular area, an under-crossing can be related to an over-crossing as shown in \autoref{fg:cross}.
\vspace{-.20cm}
In terms of $W(L_+)$ and  $W(L_-)$, this deformation can be written as
\al{
\expval{W(L_+)}=\expval{W(L_-)}+\expval{\dots U(1,x)i\Sigma^{\mu\nu}F^a_{\mu\nu}t_R^a\ U(x,2)\dots U(3,4)\dots}.
}
Once again, we can replace $F_{\mu\nu}$ using \eqref{eq:F-CSeqmot}. Then the functional derivative will act on both $1\rightarrow2$ and $3\rightarrow4$ paths in the last term. For the path $1\rightarrow2$, $dy^\alpha$ is on the plane defined by $\Sigma^{\mu\nu}$ which leads to the option \eqref{eq:deltW1}. Then, we get 
\al{
\label{eq:W+-0}
\expval{W(L_+)}=\expval{W(L_-)}-i\frac{4\pi}{k}\underset{a}\sum \expval{\dots U(1,x)\ t_R^a\ U(x,2)\dots  U(3,x)\ t_R'^a\ U(x,4)\dots}.
}
To get a recursive skein-like relation, the last term needs to be related to $W(L_0)$. The simplest way of doing this is choosing $t_R=t_R'=t_F$, where $t_F$ is in fundamental representation. Now, we can use the Fierz identity for SU(N),
\al{
\underset{a}\sum t^a_{ij}t^a_{kl}=\frac{1}{2}\delta_{il}\delta_{jk}-\frac{1}{2N}\delta_{ij}\delta_{kl}.
}
Then, \eqref{eq:W+-0} becomes
\al{
\expval{W(L_+)}=\pa{ 1+i\frac{2\pi}{k}\frac{1}{N} }\expval{W(L_-)}-i\frac{2\pi}{k}\expval{W(L_-)}.
}
This recursive relation can be solved without introducing any new restrictions on $k$. But to have the skein relation
\al{
\label{eq:skeinw0}
\beta \langle W(L_+) \rangle -\beta^{-1} \langle W(L_-)\rangle =z(\beta) \langle W(L_0) \rangle,
}
$k$ needs to be a large integer, with the following definitions:
\al{
\label{eq:betaz}
\beta=1-i\frac{2\pi}{k}\frac{1}{2N}+\mc{O}\pa{\frac{1}{k^2}}~\text{and}~ z=-i\frac{2\pi}{k}+\mc{O}\pa{\frac{1}{k^2}}.
}
Equation \eqref{eq:skeinw0}, together with \eqref{eq:Walpha}, show that Chern-Simons theory Wilson loop expectation values satisfy \eqref{eq:genskein}.
Equations \eqref{eq:betaz} relate the skein parameters to the physical constants of Chern-Simons theory. With these solutions, we have shown that Wilson loop expectation values of Chern-Simons theory are link invariants that can be obtained by solving \eqref{eq:skeinw0}, with the knot diagrams given in \autoref{fg:skein}.

\section{Geometric Quantization}\label{sec:geoquan}

In geometric quantization\footnote{In this section, we will mostly follow refs. \citen{Nair:2005iw, hall2013}}, one starts with taking the phase space of a classical theory and constructs a pre-quantum Hilbert space. This step prepares the phase space for quantization and it is not necessary in other quantization methods, like path integral formulation or canonical quantization using creation and annihilation operators. In the pre-quantum Hilbert space, wave-functions depend on all phase space variables; while in the quantum theory, they should depend on half of them that commute with each other. This feature makes the pre-quantum Hilbert space too large to describe a real quantum system. Thus, a reduction of the pre-quantum Hilbert space becomes necessary.

Quantization of a classical theory replaces the algebra of Poisson brackets with the algebra of commutation rules for quantum operators. The quantization process should lead to a Hilbert space that provides an irreducible representation of the operator algebra. This is done by choosing a subspace of the pre-quantum Hilbert space, which forces the wave-functions to depend on half of the phase space variables that commute. This ensures that quantum wave-functions are not simultaneous eigenstates of non-commuting observables. There are three general ways of doing this Hilbert space reduction: choosing the position Hilbert space, the momentum Hilbert space or the Segal-Bargmann space. This reduction procedure is called choosing a \emph{polarization} for the wave-functions. This is done by choosing a direction in the phase space that leaves the wave-function constant. It is called \emph{polarization} due to its analogous nature compared to the polarization of electromagnetic waves.

\subsection{Phase Space Geometry}\label{ssec:phasespace}

Before we get into the details about polarizations, we shall discuss how to obtain the geometry of the phase space from a given Lagrangian. Phase space is a smooth even dimensional manifold that will be denoted as $\mc{M}$. We will start with varying the action $S=\int d^4x\ \mc{L}$ with a general Lagrangian of a spin 1 gauge field. The variation of the Lagrangian is
\vspace{0.20cm}
\al{
\label{eq:varlag}
\bs
\delta \mc{L}=&\frac{\partial \mc{L}}{\partial A_\mu}\delta A_\mu+\frac{\partial \mc{L}}{\partial (\partial_\mu A_\nu)}\partial_\mu \delta A_\nu\\
=&\pb{\frac{\partial \mc{L}}{\partial A_\mu}-\partial_\nu \frac{\partial \mc{L}}{\partial (\partial_\nu A_\mu)}}\delta A_\mu+\partial_\nu \pa{\frac{\partial \mc{L}}{\partial (\partial_\nu A_\mu)}\delta A_\mu}.
\es
}
\vskip0.35cm\noindent
It is standard to assume that either $\delta A_\mu$ or $\frac{\partial \mc{L}}{\partial (\partial_\nu A_\mu)}$ vanishes at the spatial boundary $\partial V$; but assuming that it vanishes at time boundary $t_i$ and $t_f$ depends on the quantization method. In canonical or geometric quantization, this second assumption is not forced. Instead, the phase space geometry is obtained from the boundary term of \eqref{eq:varlag}, since it is a function of only the phase space variables $A_\mu$ and their conjugate momenta $\Pi^\mu$.

The boundary term from the action is
\al{
\int_{t_i}^{t_f} dt\int_V d^3x \frac{\partial \mc{L}}{\partial (\partial_0 A_\mu)}\delta A_\mu.
}
The integrand of the time integral is called the symplectic potential. Since $\Pi^\mu =\frac{\partial \mc{L}}{\partial (\partial_0 A_\mu)}$ we can write the symplectic potential as
\al{
\mathscr{A}=\int d^3x\  \Pi^\mu \delta A_\mu.
}
Here $\delta$ is the exterior differentiation in the phase space, thus it satisfies the Poincare lemma $\delta^2=0$. The symplectic two-form is given by $\Omega=\delta \mathscr{A}$ thus,
\al{
\Omega=\int d^3x\  \delta \Pi^\mu \delta A_\mu.
}
$\Omega$ plays the role of a metric and it defines the geometry of the phase space. The Poisson brackets can be obtained from $\Omega^{-1}$ as
\al{
\{f,g\}=\Omega^{\mu\nu}\partial_\mu f  \partial_\nu g
}
where $f$ and $g$ are functions that live on $\mc{M}$.

Symplectic two-form is invariant under canonical transformations. But symplectic potential transforms as $\mathscr{A'} \rightarrow \mathscr{A}+\delta \Lambda$, where $\Lambda$ is some function of the phase space variables. In some sense $\mathscr{A}$ transforms like a $U(1)$ gauge field.

\subsection{Pre-quantization}

For functions $f$ and $g$ that live on $\mc{M}$, we introduce the corresponding pre-quantum operators $Q(f)$ and $Q(g)$ by
\al{
Q_{pre}(\{f,g\})=\frac{1}{i\hbar}[Q_{pre}(f), Q_{pre}(g)].
}
This equation is solved by
\al{
\label{eq:preqopr}
Q_{pre}(f)=i\hbar\pa{X_f-\frac{i}{\hbar}\mathscr{A}_j X^j_f}+f
}
where $\mathscr{A}_j$ are components of the symplectic potential and $X_f$ is a Hamiltonian vector field given by
\al{
X_f=\frac{\partial f}{\partial x_i}\frac{\partial }{\partial p_i}-\frac{\partial f}{\partial p_i}\frac{\partial }{\partial x_i}.
}
To describe a single point particle, we can choose $f=x_i$ and $g=p_i$. Then the components of vector $X$ are
\al{
X_{x_i}=\frac{\partial}{\partial p_i} \ ~\text{and}~\ X_{p_i}=-\frac{\partial}{\partial x_i}.
}
Using $\mathscr{A}=p_i dx_i$, we get
\al{
\label{eq:Qpre}
Q_{pre}(x_i)=x_i+i\hbar\frac{\partial}{\partial p_i} \ ~\text{and}~\ Q_{pre}(p_i)=-i\hbar\frac{\partial}{\partial x_i}.
}
These operators satisfy $[Q_{pre}(x_i),Q_{pre}(p_i)]=i\hbar$; thus, \eqref{eq:Qpre} is a representation. But it is not irreducible. This problem comes from pre-quantum Hilbert space being too large. The solution is to choose a subspace of the pre-quantum Hilbert space and this is done by \emph{polarizing} the wave-function. The pre-quantum wave-function is a function of all phase space variables, i.e. $\Phi=\Phi(x_i,p_i)$. Polarization forces the quantum wave-function to depend on half of the phase space variables that commute. There are three general ways of doing this: choosing the position Hilbert space where the quantum wave-function $\psi=\psi(x_i)$, choosing the momentum Hilbert space where $\psi=\psi(p_i)$, or choosing the Segal-Bargmann space where $\psi=\psi(z)$ and $z=x+ip$. The first two options are obtained with real polarizations and the last one is with holomorphic polarization.

The polarization condition is defined by 
\al{
\mc{D}_i\Phi=0,
}
where $\mc{D}_i$ is a covariant derivative given by $\mc{D}_i=\partial_i-i\mathscr{A}_i$. Then, the curvature is given by the commutator $[\mc{D}_i,\mc{D}_j]=i\Omega_{ij}$. 

For a polarization to be well defined, the wave-function has to be square integrable. Otherwise a well define inner product cannot be defined. The pre-quantum inner product is given by
\al{
\label{eq:preqin}
\expval{1|2}=\int d\sigma(\mc{M})\ \Phi_1^*\Phi_2^{ },
}
where integration is done over all phase space variables. Under canonical transformation $\mathscr{A'} \rightarrow \mathscr{A}+\delta \Lambda$ the pre-quantum wave functional transforms like a matter field, as $\Phi' \rightarrow e^{i\Lambda}\Phi$. Thus, \eqref{eq:preqin} is invariant under canonical transformations, as one would expect. 

\subsection{Real Polarizations}

For a point particle in one dimension, symplectic two-form is given by,
\al{
\Omega=dp\wedge dx
}
and up to a canonical transformation, the symplectic potential is
\al{
\mathscr{A}=pdx.
}
As we have shown, the pre-quantum operators for position $\mc{Q}_{pre}(x)$ and momentum $\mc{Q}_{pre}(p)$ should satisfy
\al{
[\mc{Q}_{pre}(x),\mc{Q}_{pre}(p)]=i\hbar.
}
One representation that satisfies this commutator is given by
\al{
\mc{Q}_{pre}(x)=i\hbar\frac{\partial}{\partial p}+x\ ~\text{and}~\ \mc{Q}_{pre}(p)=-i\hbar\frac{\partial}{\partial x}.
}

In $(x,p)$ coordinates, there are two polarization choices: 
\al{
\mc{D}_p\Phi(x,p)=0\ ~\text{or}~\ \mc{D}_x\Phi(x,p)=0.
} 
For $\mathscr{A}=pdx$, $\mc{D}_p=\partial_p$ and $\mc{D}_x=\partial_x-ip$. Choosing the first polarization condition gives $\Phi(x,p)=\psi(x)$, where $\psi$ is the quantum wave-function. For this case, the pre-quantum operators are
\al{
\mc{Q}_{pre}(x)=x\ ~\text{and}~\ \mc{Q}_{pre}(p)=-i\hbar\frac{\partial}{\partial x}.
}
This is an irreducible representation. Choosing the second condition leads to  $\Phi(x,p)=\psi(p)e^{ipx}$. One can always do a canonical transformation, such as 
\al{
\mathscr{A}\rightarrow\mathscr{A'}=\mathscr{A}-d(xp)=-xdp,
}
to obtain $\mc{D}_p=\partial_p+ix$ and $\mc{D}_x=\partial_x$. Then, choosing $\mc{D}_x\Phi=0$ gives $\Phi(x,p)=\psi(p)$ and quantum operators for this case are
\al{
\mc{Q}_{pre}(x)=i\hbar\frac{\partial}{\partial p}\ ~\text{and}~\ \mc{Q}_{pre}(p)=p.
}
For all of these cases, $\psi$ will depend on either $x$ or $p$. 

In real polarizations, the pre-quantum inner product
\al{
\expval{\psi|\psi}=\int dx\ dp\ \psi^*(x)\psi(x)
}
is not finite. Thus, it cannot be used as the quantum inner product. The irrelevant degrees of freedom should not be integrated over. If the chosen subspace is the position Hilbert space, then integration should be over just $dx_i$ (or $dp_i$ for the momentum Hilbert space) as,
\al{
\langle \psi|\psi \rangle = \int dx\ \psi^*(x)\psi(x)\ ~\text{or}~\ \langle \psi|\psi \rangle = \int dp\ \psi^*(p)\psi(p).
}
It is well known that these two cases are physically equivalent. The equivalence can be shown using the fact that $\psi(x)$ and $\psi(p)$ are Fourier transforms of each other. But just not integrating over $dp$ in the position Hilbert space (or $dx$ in the momentum Hilbert space) is not the proper way of defining a finite inner product. The correct way of doing this is called the half-form quantization.

\subsection{Half-Form Quantization}

A half-form is the square root of a one-form, defined as
\al{
\sqrt{dx}\otimes\sqrt{dx}=dx.
}
To define a finite inner product, instead of using $\psi$ and integral measure as separate objects, a composite object $s=\psi\otimes\sqrt{dx}$ needs to be defined as a polarized section. Then the inner product is given by
\al{
||s||^2=\int |\psi|^2\ dx.
}
This procedure makes sure that the integration is done over only the relevant degree of freedom and the inner product is finite. 

In half-form quantization, quantum operators act on $s$, not just $\psi$. An operator $Q(f)$ on the half-form space acting on $s$ is given by\cite{hall2013}
\al{
\label{eq:half-formQ}
Q(f)s=\big( Q_{pre}(f)\psi \big) \otimes \sqrt{dz} -i\hbar\ \psi \otimes \mc{L}_{X_f}\sqrt{dz}
}
where $\mc{L}$ is a Lie derivative. The second term in \eqref{eq:half-formQ} is called the ``metaplectic correction".

\subsection{Complex Polarizations}

When the phase space is K\"ahler, it is advantageous to work with complex coordinates. Introducing the local complex coordinates $z^a, \bar{z}^a$, we can write
\al{
\Omega=\frac{1}{2}\Omega_{a\bar{a}}dz^a\wedge d\bar{z}^{\bar{a}}.
}
The metric is given by
\al{
g_{a\bar{b}}=\partial_a \partial_{\bar{b}} K,
}
where K is a real function,n called the K\"ahler potential. Metric components and symplectic two-form components are related by $\Omega_{a\bar{b}}=-\Omega_{\bar{b}a}=ig_{a\bar{b}}$. The symplectic two-form can also be written as
\al{
\Omega = i\partial \bar{\partial} K,
}
where $\partial = dz^a \wedge \frac{\partial}{\partial z^a}$ and $\bar{\partial} = d\bar{z}^{\bar{a}} \wedge \frac{\partial}{\partial \bar{z}^{\bar{a}}}$.
 
With the K\"ahler potential, two covariant derivatives can be defined as
\al{
\label{eq:covdev}
\mc{D}_a=\partial_a - i\mathscr{A}_a~\text{and}~
\mc{D}_{\bar{a}}=\partial_{\bar{a}} - i\mathscr{A}_{\bar{a}},
}
where the connections are given by
\al{
\label{eq:symppot}
\mathscr{A}_a=-\frac{i}{2}\partial_a K~\text{and}~\mathscr{A}_{\bar{a}}=\frac{i}{2}\partial_{\bar{a}}K.
}
These are the components of the symplectic potential that is given by $\Omega=d\mathscr{A}$. Thus, the commutator of the covariant derivatives in \eqref{eq:covdev} satisfy $[\mc{D}_a,\mc{D}_{\bar{a}}]=i\Omega_{a \bar{a}}$.

With the covariant derivatives defined above, two polarization conditions can be written: $\mc{D}_{\bar{a}}\Phi=0$ and  $\mc{D}_a\Phi=0$. The first one
\al{
\mc{D}_{\bar{a}}\Phi=\pa{\partial_{\bar{a}}+\frac{1}{2}\partial_{\bar{a}}K}\Phi=0
}
is solved by
\al{
\Phi(z^a,\bar{z}^a)=e^{-\frac{1}{2}K(z^a,\bar{z}^a)}\Psi(z^a)
}
where $\Psi$ is holomorphic in $z^a$. In holomorphic polarizations, the pre-quantum inner product can be retained at the quantum level as,
\al{
\langle \Phi_1|\Phi_2 \rangle= \int d\sigma(\mc{M})\ \Phi_1^*\Phi_2\ \longrightarrow\ \langle\psi_1|\psi_2 \rangle=\int d\sigma(\mc{M})\ e^{-K}\psi_1^*\psi_2.
}
This inner product is well behaved only if $K$ is positively defined. For $K<0$, $\mc{D}_a\Phi=0$ has to be chosen. Clearly, only one of the two polarizations will give a well behaved inner product. Hence, the appropriate one must be chosen. 

One important point that distinguishes the holomorphic polarization from real polarizations is the integral measure. In real polarizations, integration is done only over the polarized section. However, in holomorphic polarizations, the integration is done over the whole phase space volume. $\psi$ might just depend on $z$, but $\psi^*$ will have $\bar{z}$ dependence and $K$ always depends on both $z$ and $\bar{z}$. Thus, integration over both $z$ and $\bar{z}$ is necessary. This feature seem to make half-from quantization unnecessary for holomorphic polarizations.  But there still are advantages in using half-forms, even though the inner product is already finite without using them.

The most well known example for holomorphic quantization is the coherent state quantization of a simple harmonic oscillator. The Hamiltonian for this system is given by
\al{
H=\frac{1}{2m}\pa{p^2+(m\omega x)^2}=\frac{1}{2m}\pa{p^2+y^2},
}
where $y=m\omega x$. The symplectic potential is
\al{
\mathscr{A}=\frac{1}{2}(pdx-xdp)=\frac{1}{2m\omega}(pdy-ydp).
}
Using \eqref{eq:preqopr} with $f=H$, we get
\al{
Q_{pre}(H)=i\hbar\pa{y\frac{\partial}{\partial p}-p\frac{\partial}{\partial y}}.
}
Now, we switch to complex coordinates
\al{
z=\frac{1}{m\omega}(y-ip)\ ~\text{and}~\ \bar{z}=\frac{1}{m\omega}(y+ip).
}
Then, choosing the polarization condition
\al{
D_{\bar{z}}\Phi(z,\bar{z})=\pa{\partial_{\bar{z}}+\frac{m\omega}{2\hbar}z}\Phi(z,\bar{z})=0
}
gives
\al{
\Phi(z,\bar{z})=\psi(z)e^{-\frac{m\omega}{2\hbar}z\bar{z}}
}
and $K=\frac{m\omega}{\hbar}z\bar{z}$ is the K\"ahler potential. Now, we can write
\al{
Q_{pre}(H)\ \psi=\hbar\pa{y\frac{\partial}{\partial p}-p\frac{\partial}{\partial y}}\psi\pa{\frac{1}{m\omega}(y-ip)}=\hbar\omega z \frac{d\psi}{dz}.
}
$\psi$ is in the form\cite{hall2013} of $\psi=\sum \limits_n a_n z^n$. Thus, $z \frac{d\psi}{dz}=n\psi$ and this gives the spectrum
\al{
E_n=\hbar\omega n.
}
This differs from the \emph{correct} spectrum of the harmonic oscillator by $\hbar \omega /2$. To obtain the correct spectrum, the system needs to be quantized with half-forms. 

For $f=H$, the Lie derivative acting on $\sqrt{dz}$ in the metaplectic correction term is given by\cite{hall2013}
\al{
\mc{L}_{X_H}\sqrt{dz}=\frac{i\omega}{2}\sqrt{dz}.
}
Then using \eqref{eq:half-formQ}, we can write
\al{
Q(H)\ s=\hbar \omega n\ \psi \otimes \sqrt{dz} + \frac{\hbar\omega}{2}\ \psi \otimes \sqrt{dz}.
}
From this equation, it can be seen that the missing $\hbar \omega /2$ in the spectrum appears as a metaplectic correction, giving the \emph{correct} spectrum
\al{
E_n=\hbar\omega \pa{n+\frac{1}{2}}.
}
Although holomorphic polarizations lead to a well defined quantization even without the help of half-forms, in some cases like this, it can be advantageous to do half-form quantization. However, we will not use half-forms for Chern-Simons theory, since there seems to be no natural way of doing so.

\subsection{Geometric Quantization of Gauge Theories}\label{eq:geoquangauge}

In quantum mechanics, solving Schr\"odinger's equation is sufficient to find the wave-function. However, in a gauge field theory, Gauss' law also needs to be solved in addition to Schr\"odinger's equation. 

Gauss' law operator is the generator of infinitesimal gauge transformations of a given gauge theory. In $SU(N)$ gauge theories,  the gauge field transforms as
\al{
A_\mu \rightarrow A_\mu^g = g A_\mu g^{-1} - (\partial_\mu g) g^{-1}
}
where $g=exp(-it^a \theta^a)$. For $\theta \ll 1$, the infinitesimal transformation ($g\approx 1-it^a\theta^a$) can be written as
\al{
A^g_\mu = A_\mu + i t^a (D_\mu \theta)^a.
}
Infinitesimal transformations are generated by the vector field
\al{
\label{eq:xi}
\xi = \int d^nx\ \delta A_\mu^a \frac{\delta}{\delta A^a_\mu}= - \int d^nx\ (D_\mu \theta)^a \frac{\delta}{\delta A^a_\mu},
}
where $n$ is the dimension of the space of gauge fields. To find the Gauss' law operator, we write the interior contraction of the symplectic two-form with the vector field,
\al{
i_\xi \Omega = \xi^\mu \Omega_{\mu\nu} dq^\nu,
}
where $q^\nu$ is a phase space coordinate. Generally, for transformations in the phase space, the interior contraction satisfies
\al{
\label{eq:intcont}
i_\xi \Omega = - \delta f,
}
where $f$ is some function of the phase space variables. For infinitesimal gauge transformations, $f=\int G^a\theta^a$ and then \eqref{eq:intcont} becomes
\al{
\label{eq:intcont2}
i_\xi \Omega = - \delta \int d^nx\ G^a \theta^a.
}
Here $G^a$ is the Gauss' law operator and the Gauss' law is given by
\al{
\label{eq:gauss0}
G^a \psi=0.
}
$G^a=0$ also appears as one of the equations of motion.

At least for the gauge theories that we are interested in, such as CS, YM or TMYM, the wave-functional can be factorized in the form $\psi=\phi\chi$. Here, $\phi$ is the part that satisfies Gauss' law constraint $G^a\Psi=0$ and $\chi$ is the gauge invariant part that is necessary for $\psi$ to satisfy the Schr\"odinger's equation $\mc{H}\psi=\mc{E}\psi$. To find $\phi$, the standard technique is to make an infinitesimal gauge transformation on $\psi$, then force the Gauss' law to obtain a condition that is usually solved by some Wess-Zumino-Witten(WZW) action. Once $\phi$ is known, then the Schr\"odinger's equation can be tackled to find $\chi$. In 2+1 dimensional gauge theories, $\chi$ is where the scale dependence is hidden.
\chapter{Geometric Quantization of Chern-Simons Theory}\label{ch:cs}

In this chapter, we will review the geometric quantization of non-Abelian Chern-Simons(CS) theory, mostly following Bos and Nair's work \cite{Bos:1989kn,Nair:2005iw}\footnote{Another comprehensive discussion on this subject can be found in ref. \citen{axelrod}}.

The CS action is given by
\al{
S_{CS}=-\frac{k}{4\pi} \int \limits_{\Sigma\times[t_i,t_f]} {d^3x}\ \epsilon^{\mu\nu\alpha}\ Tr \pa{A_\mu \partial_\nu A_\alpha + \frac{2}{3}A_\mu A_\nu A_\alpha},
}
where $\Sigma$ is an orientable two dimensional surface. This action is classically not gauge invariant, but in the quantum theory it can be made gauge invariant by restricting $k$ to be an integer, as discussed in \autoref{sec:csint}.
The equations of motion for this theory are
\al{
F_{\mu\nu}=\partial_\mu A_\nu-\partial_\nu A_\mu+[A_\mu,A_\nu] =0.
}

In the temporal gauge $A_0=0$ with complex coordinates $A_z=\frac{1}{2}(A_1+ iA_2)$ and $A_{\bar{z}}=\frac{1}{2}(A_1- iA_2)$, the CS action becomes
\al{
S_{CS}=-\frac{ik}{2\pi}\int{dt d\mu_\Sigma }\ Tr(A_{\bar z}\partial_0 A_z - A_{z}\partial_0 A_{\bar z} ).
}
The equations of motion ($F_{\mu\nu}=0$) in this gauge makes $A_z$ and $A_{\bar{z}}$ time independent along with constraining $F_{z\bar{z}}=0$. A very important feature is that the conjugate momenta are given by the gauge fields,
\al{
\Pi^{a z}=\frac{ik}{4\pi} A^a_{\bar{z}}~\ \text{and}~\ \Pi^{a \bar{z}}=-\frac{ik}{4\pi}  A^a_z.
}
Later, we will see a similar behavior in TMYM theory which is crucial for our work.

Following the method outlined in \autoref{ssec:phasespace}, we write the symplectic potential as
\al{
\mathscr{A}=\frac{ik}{4\pi}\int \limits_{\Sigma} (A^a_{\bar{z}}\delta A^a_z-A^a_z\delta A^a_{\bar{z}}).
}
The symplectic two-form of CS theory is given by $\Omega=\delta \mathscr{A}$,
\al{
\label{eq:omegacs}
\Omega=\frac{ik}{2\pi}\int \limits_{\Sigma}\delta A^a_{\bar{z}}\delta A^a_z.
}
This phase space is K\"ahler and the K\"ahler potential is given by $\Omega=i\partial\bar{\partial}K$, where $\partial=\delta A^a_z \wedge \frac{\partial}{\partial A^a_z}$ and $\bar{\partial}=\delta A^a_{\bar{z}} \wedge \frac{\partial}{\partial A^a_{\bar{z}}}$; hence,
\al{
\label{eq:kahlercs}
K=\frac{k}{2\pi} \int \limits_{\Sigma} A^a_{\bar{z}}A^a_z.
}

As we discussed in \autoref{sec:WZW}, in simply connected spaces it is possible to parametrize the gauge fields as 
\al{
\label{eq:par1}
A_{\bar z}=-\partial_{\bar z}UU^{-1}~\text{and}~A_z=U^{\dagger-1}\partial_z U^{\dagger}.
}
Here, U is a complex SL(N,$\mathbb{C}$) matrix which transforms like $U^g=gU$  where $g\in\mc{G}$ and $\mc{G}$ is the gauge group. We will continue with taking $\mc{G}=SU(N)$. $U$ is given by
\al{
\label{eq:U}
U(x,0,C)=\mc{P}exp\pa{-\underset{C}{\ \ \int_0^x}(A_{\bar{z}}d\bar{z}+\mc{A}_zdz)},
}
where $\mc{A}_z$ satisfies $\partial_z A_{\bar{z}}-\partial_{\bar{z}}\mc{A}_z+[\mc{A}_z,A_{\bar{z}}]=0$ and this flatness condition makes $U$ invariant under small deformations of the path $C$ on $\Sigma$.  From \eqref{eq:U}, it follows that
\al{
\label{eq:scriptA}
\mc{A}_z=-\partial_z U U^{-1} ~\text{and}~ \mc{A}_{\bar{z}}= U^{\dagger-1}\partial_{\bar{z}} U^{\dagger}.
}

\section{Gauss' Law}\label{sec:gausscs}

Infinitesimal gauge transformations ($g\approx 1-it^a\varepsilon^a$)  are generated by the vector field given by \eqref{eq:xi}. In complex coordinates, it is
\al{
\xi=-\int \limits_{\Sigma} \pb{(D_z\varepsilon)^a\frac{\delta}{\delta A^a_z}+(D_{\bar{z}}\varepsilon)^a\frac{\delta}{\delta A^a_{\bar{z}}}}.
}
To find the Gauss' law, we look at the interior contraction
\vspace{0.20cm}
\al{
\bs
i_\xi \Omega =& -\frac{ik}{2\pi} \int \limits_{\Sigma} \varepsilon^a \pb{ -(D_z\delta A_{\bar{z}})^a+(D_{\bar{z}}\delta A_z)^a }\\
=& -\delta \pa{\frac{ik}{2\pi} \int \limits_{\Sigma} F^a_{z\bar{z}}\varepsilon^a}.
\es
}
\vskip0.35cm\noindent
Then using \eqref{eq:intcont2}, we can write the Gauss' law operator for CS theory as
\al{
G^a=\frac{ik}{2\pi} F^a_{z\bar{z}}.
}


\section{The Wave-Functional}

We choose the holomorphic polarization, which makes the quantum wave-functional $\psi$ only $A_{\bar{z}}$ dependent. The pre-quantum and quantum wave-functionals are related by $\Phi[A_z,A_{\bar{z}}]=e^{-\frac{1}{2}K}\psi[A_{\bar{z}}]$, where $K$ is the K\"ahler potential given by \eqref{eq:kahlercs}. Since the phase space is K\"ahler, the pre-quantum inner product can be retained at the quantum level, as
\al{
\langle 1|2 \rangle =\int d\mu(\mc{M})\ \Phi_1^*\Phi_2 ~\rightarrow~ \int d\mu(\mc{M})\ e^{-K}\psi_1^*\psi_2
}
where $d\mu(\mc{M)}$ is the Liouville measure defined by the symplectic two-form.

Upon quantization we can write,
\al{
\label{eq:delta}
A^a_z\psi[A^a_{\bar{z}}]=\frac{2\pi}{k} \frac{\delta}{\delta A^a_{\bar{z}}} \psi[A^a_{\bar{z}}].
}

Since no currents are present, the wave-functional must satisfy $F_{z\bar{z}}\ \psi[A_{\bar{z}}]=0$ which is the Gauss' law of CS theory. We then make an infinitesimal gauge transformation on the wave-functional $\psi$ with parameter $\varepsilon$,
\vspace{0.20cm}
\al{
\bs
\delta_\varepsilon \psi[A_{\bar{z}}]=&\int d^2z\ \delta_\varepsilon A^a_{\bar{z}}\ \frac{\delta\psi}{\delta A^a_{\bar{z}}} \\
=& \int d^2z\ \varepsilon^a \pa{ \partial_{\bar{z}} \frac{\delta}{\delta A^a_{\bar{z}}} + i f^{abc} A^b_{\bar{z}}  \frac{\delta}{\delta A^c_{\bar{z}}} }\psi\\
=& -\frac{k}{2\pi} \int d^2z\ \varepsilon^a (F^a_{z\bar{z}} - \partial_z A^a_{\bar{z}}) \psi.
\es
}
\vskip0.35cm\noindent
Then applying the Gauss' Law constraint $F_{z\bar{z}}\ \psi[A_{\bar{z}}]=0$ gives
\al{
\label{eq:infg}
\delta_\varepsilon \psi[A_{\bar{z}}]= \frac{k}{2\pi} \int d^2z\ \varepsilon^a  \partial_z A^a_{\bar{z}} \ \psi[A_{\bar{z}}].
}
From \eqref{eq:wzwtransform}, we know that this condition is solved by writing
\al{
\label{eq:cswf}
\psi[A_{\bar{z}}]=exp\big( kS_{WZW}(U)\big).
}
In general, the wave-functional in \eqref{eq:cswf} can have a gauge invariant factor $\chi$ which can be found by solving the Schrodinger's equation $\mc{H}\psi=\mc{E}\psi$. But since the CS Hamiltonian for ground state is zero in the temporal gauge, we take $\chi=1$. $\chi$ is where the scale dependence is hidden. For a topological theory like CS, a constant $\chi$ is expected.


\section{The Measure}\label{sec:csmeasure}
Using the symplectic two-form of CS theory \eqref{eq:omegacs}, we can write the metric on $\mathfrak{A}$, the space of gauge potentials\cite{Karabali1996135}, as
\vspace{0.20cm}
\al{
\bs
ds^2_{\mathfrak{A}}=&\int d^2x\ \delta A^a_i \delta A^a_i=-8\int Tr(\delta A_{\bar{z}} \delta A_z)\\
=& 8 \int Tr[D_{\bar{z}}(\delta U U^{-1})D_z(U^{\dagger -1}\delta U^{\dagger})].
\es
}
\vskip0.35cm\noindent
Our goal is to write the volume element of this space in terms of the volume of $SL(N,\mathbb{C})$, which has the Cartan-Killing metric
\al{
ds^2_{SL(N,\mathbb{C})}=8\int Tr[(\delta U U^{-1})(U^{\dagger -1}\delta U^{\dagger})].
}
The volumes of these two spaces are related by
\al{
d\mu(\mathfrak{A})=det(D_{\bar{z}}D_z)\ d\mu(U,U^{\dagger}).
}
This measure is not gauge invariant. To make it invariant, we use the $SU(N)$ gauge invariant matrix $H=U^{\dagger}U$ that we defined in \autoref{sec:WZW}, where $H \in SL(N,\mathbb{C})/SU(N)$. Now, we can write
\al{
\label{eq:CSmeasure}
d\mu(\mathfrak{A})=det(D_{\bar{z}}D_z)\ d\mu(H).
}
As shown in \autoref{sec:WZW}, the determinant is
\al{
det(D_{\bar{z}}D_z)=constant \times e^{2c_AS_{WZW}(H)},
}
where $c_A$ is the quadratic Casimir in the adjoint representation given by $c_A\delta^{ab}=f^{amn}f^{bmn}$.

Now that we have the measure and the wave functional, we can write the inner product
\al{
\langle \psi_1|\psi_2 \rangle=\int d\mu(\mathfrak{A})\ e^{-K}\ \psi^*_1\psi^{ }_2.
}
Using the Polyakov-Wiegmann identity we get,
\al{
e^{-K}\psi^*\psi=e^{kS_{WZW}(H)}.
}
Then, the inner product for $\psi$ becomes
\al{
\label{eq:csinnerproduct}
\langle \psi|\psi \rangle=\int d\mu(H)\ e^{(2c_A+k)S_{WZW}(H)}.
}
\boldmath 
\section{The $ \Sigma = S^1 \times S^1 $ Case}
\unboldmath
If the space is not simply connected, the parametrization we used in \eqref{eq:par1} needs modification. On a torus, as discussed in \autoref{sec:WZW}, the correct parametrization is
\al{
A_{\bar{z}}=-\partial_{\bar{z}}UU^{-1}+Ui\pi(Im\ \tau)^{-1}aU^{-1},
}
where $\tau$ is the modular parameter of the torus and $a$ is a constant gauge field. The new term can be absorbed in a matrix as $V=U\ exp[i\pi(Im\ \tau)^{-1}(z-\bar{z})a]$ and then $A_{\bar{z}}$ can be once again parametrized in the form $-\partial_{\bar{z}}VV^{-1}$. But this gives rise to a new factor in the wave-functional which depends on $a$. Now the wave-functional is
\al{
\label{eq:toruspsi}
\psi[A_{\bar{z}}]=exp\big( kS_{WZW}(V)\big) \Upsilon(a).
}
Finding this new factor is not straightforward and we will not review its calculation here, but the result can be found in ref. \citen{Bos:1989kn}. 


\section{Wilson Loops in Chern-Simons Theory}\label{sec:wilcs}

\begin{figure}[H]
\centering
\includegraphics[scale=0.53]{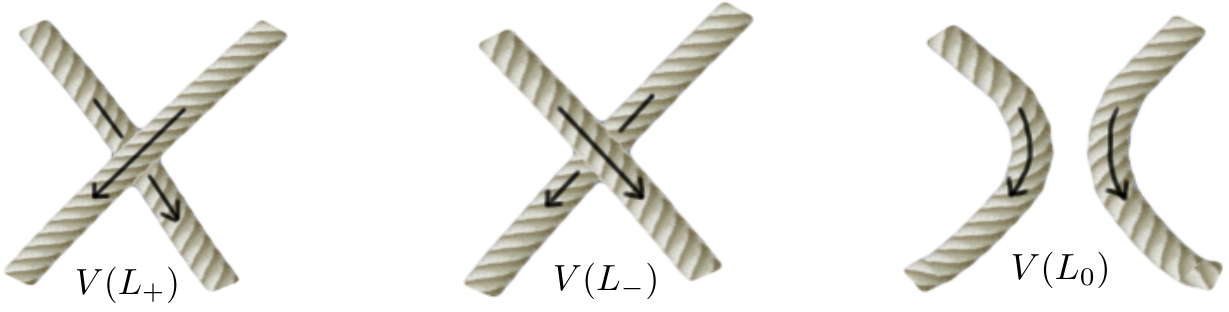}
\caption{Knot diagrams}
\label{fg:skeinch3}
\end{figure}

The Wilson loop operator for representation $R$ and path $C$ is given by
\al{
\label{eq:wlsn}
W_R(C)=Tr_R\ \mc{P}\ e^{-\oint \limits_c A_\mu dx^\mu}.
}
As shown in \autoref{sec:knot}, in CS theory, the expectation value of this operator can be calculated directly from skein relations without any field theory calculation. Up to some approximation, a generalized skein relation can be obtained\cite{CottaRamusino:1989rf} for WLEVs in the fundamental representation, as
\al{
\label{eq:skeinw}
\beta \langle W_{L_+} \rangle -\beta^{-1}\langle W_{L_-}\rangle=z(\beta)\langle W_{L_0}\rangle
}
where
\al{
\beta=1-i\frac{2\pi}{k}\frac{1}{2N}+\mc{O}\pa{\frac{1}{k^2}}~\text{and}~ z=-i\frac{2\pi}{k}+\mc{O}\pa{\frac{1}{k^2}}
}
and the knot diagrams are shown in \autoref{fg:skeinch3}.
\vspace{-.20cm}

In the temporal gauge with complex coordinates and in representation $R$, \eqref{eq:wlsn} becomes
\al{
W_R(C)=Tr_R\ \mc{P}\ e^{-\oint \limits_c (A_zdz+A_{\bar{z}}d\bar{z})}.
}
Since $A_z$ is the derivative with respect to $A_{\bar{z}}$ and it acts on everything on its right, expanding this path ordered exponential leads to a very difficult calculation. To avoid this, instead of using the usual definition of the Wilson loop, we would like to use a \emph{Wilson loop-like} observable defined as
\al{
\mc{W}_R(C)=Tr_R\ \mc{P}\ e^{-\oint \limits_c (\mc{A}_zdz+A_{\bar{z}}d\bar{z})}=Tr_R\ U(x,x,C),
}
where $\mc{A}_z$ is given by $\partial_z A_{\bar{z}}-\partial_{\bar{z}}\mc{A}_z+[\mc{A}_z,A_{\bar{z}}]=0$. Since $F_{z\bar{z}}=0$ from the Gauss' law, replacing $A_z$ with $\mc{A}_z$ does not change the general properties of the Wilson loop when evaluated on states that live on the constraint hyper-surface. Since $U$ is defined to be path independent, it would seem that the skein relations are trivially satisfied. However, this is not true since the path independence is only on $\Sigma$, because we are forcing flatness only on the $z\bar{z}$ component of the curvature. So, one is allowed to make small deformations in the time direction, piercing $\Sigma$ to get skein relations.

In the previous section, we have shown that the theory is given by the action $S_{WZW}(H)$, thus we can use gauge invariant WZW currents $J_{\bar{z}}=-\partial_{\bar{z}}HH^{-1}$ and $J_z=H^{-1}\partial_z H$ to write gauge invariant observables similar to Wilson loops\cite{Karabali:1998yq}. The gauge fields in $\mc{W}_R(C)$ can be written as $SL(N,\mathbb{C})$ transformed WZW currents;
\vspace{0.20cm}
\al{
\label{eq:AJ}
\bs
A_{\bar{z}}=&-\partial_{\bar{z}}UU^{-1}=U^{\dagger-1}J_{\bar{z}}U^{\dagger}+U^{\dagger-1}\partial_{\bar{z}}U^{\dagger},\\
\mc{A}_z=&-\partial_zUU^{-1}=U^{\dagger-1}J_zU^{\dagger}+U^{\dagger-1}\partial_zU^{\dagger}.
\es
}
\vskip0.35cm\noindent
Thus, we can write $\mc{W}$ in terms of H
\al{
\mc{W}_R(C,H)=Tr_R\ \mc{P}\ e^{\ \oint \limits_c (\partial_zHH^{-1}dz+\partial_{\bar{z}}HH^{-1}d\bar{z})}.
}
Since $\mc{W}$ commutes with the wave-functional, the expectation value of it can be written as
\al{
\langle \mc{W}_R(C) \rangle=\int d\mu(H)e^{(2c_a+k)S_{WZW}(H)}\mc{W}_R(C,H).
}

\chapter{Topologically Massive Yang-Mills Theory}\label{ch:tmym}

Topologically massive Yang-Mills action(TMYM) is given by
\vspace{0.20cm}
\al{
\label{eq:actiontmym}
\bs
S_{TMYM}=&S_{CS}+S_{YM}\\
=&-\frac{k}{4\pi}\int \limits_{\Sigma \times[t_i,t_f]} {d^3x }\ \epsilon^{\mu\nu\alpha}\ Tr \pa{A_\mu \partial_{\nu} A_{\alpha} + \frac{2}{3}A_\mu A_\nu A_\alpha}\\
&-\frac{k}{4\pi}\frac{1}{m}\int \limits_{\Sigma\times[t_i,t_f]} {d^3x }\ Tr\  F_{\mu\nu}F^{\mu\nu}.
\es
} 
\vskip0.35cm\noindent
Here $m$ is called the topological mass (see \autoref{sec:TMYMint}). Our definition of topological mass differs by a factor $\frac{k}{4\pi}$ compared to the literature on this theory. We made this choice so that studying different values of $k$ does not change the balance of the theory in either pure Yang-Mills(YM) or pure Chern-Simons(CS) direction. That is decided only by the value of $m$. With this choice of constants, the equations of motion are $k$ free, as
\al{
\epsilon^{\mu\alpha\beta}F_{\alpha\beta}+\frac{1}{m} D_{\nu}F^{\mu\nu}=0
}
where $D_\mu \bullet=\partial_\mu \bullet + [A_\mu,\bullet]$.
Following ref. \cite{yildirim}, we simplify the notation with defining
\al{
\label{eq:tilde0}
\tilde A_{\mu}=A_{\mu}+\frac{1}{2m}\epsilon_{\mu\alpha\beta}F^{\alpha\beta}.
}
Here $\tilde{A}$ is not a field redefinition. It is just a shorthand notation to make equations easy to compare with pure CS theory.
From \eqref{eq:tilde0}, it can be seen that $\tilde A_\mu$ transforms like a gauge field since $F^{\alpha\beta}$ is gauge covariant. For future convenience, using $\tilde A_\mu$ as a connection, we define a new covariant derivative $\tilde D_\mu \bullet=\partial_\mu \bullet + [\tilde A_\mu,\bullet]$.

Using complex coordinates and choosing the temporal gauge, the conjugate momenta are given by $\tilde{A}$,
\al{
\Pi^{a z}=\frac{ik}{4\pi} \tilde A^a_{\bar{z}}~\ \text{and}~\ \Pi^{a \bar{z}}=-\frac{ik}{4\pi}  \tilde A^a_z,
}
with
\al{
\label{eq:Az}
\tilde A_{z}=A_{z}+E_{z}~\ \text{and}~\ \tilde A_{\bar{z}}=A_{\bar{z}}+E_{\bar{z}}
}
and where
\al{
\label{eq:E}
E_{z}=\frac{i}{2m}F^{0\bar{z}} \ ~\text{and}~\  E_{\bar{z}}=-\frac{i}{2m}F^{0z}.
}
The conjugate momenta of TMYM theory transform like gauge fields and this feature gives the theory a Chern-Simons-like behavior in some sense. Thus, we will be able to borrow many of the features of CS theory in the following analysis.

The symplectic two-form for this theory is
\vspace{0.20cm}
\al{
\bs
\label{eq:omegatmym}
\Omega=&\frac{ik}{4\pi}\int \limits_{\Sigma}(\delta \tilde A^a_{\bar{z}} \delta A^a_z+\delta A^a_{\bar{z}}\delta \tilde A^a_z )\\
=&\frac{ik}{4\pi}\int \limits_{\Sigma}(2\delta A^a_{\bar{z}} \delta A^a_z +\delta E^a_{\bar{z}} \delta A^a_z+ \delta A^a_{\bar{z}}\delta E^a_z  ).
\es
}
\vskip0.35cm\noindent
From this equation, it can be seen that the TMYM phase space consists of two CS-like parts. This can be seen more clearly under a coordinate transformation. Instead of using $A_z$ and $A_{\bar{z}}$, we could use $B_z=\frac{1}{2}(A_1+i\tilde A_2)$, $C_z=\frac{1}{2}(\tilde A_1+i A_2)$ and their complex conjugates. This would allow us to write $\Omega$ in the form $\int \limits_{\Sigma}(\delta B^a_z \delta B^a_{\bar{z}} + \delta C^a_z \delta C^a_{\bar{z}})$. Thus, the phase space of TMYM theory can be written as a sum of two CS phase spaces.


\section{Gauss' Law}\label{sec:gausstmym}

In TMYM theory, infinitesimal gauge transformations are generated by the vector field
\al{
\xi=-\int \limits_{\Sigma} \pb{(D_z\varepsilon)^a\frac{\delta}{\delta A^a_z}+(D_{\bar{z}}\varepsilon)^a\frac{\delta}{\delta A^a_{\bar{z}}}+(\tilde D_z\varepsilon)^a\frac{\delta}{\delta \tilde A^a_z}+(\tilde D_{\bar{z}}\varepsilon)^a\frac{\delta}{\delta \tilde A^a_{\bar{z}}}}.
}
To find the Gauss' law, we look at the interior contraction
\vspace{0.20cm}
\al{
\bs
i_\xi \Omega =& -\frac{ik}{4\pi} \int \limits_{\Sigma} \varepsilon^a \pb{ -(D_z\delta \tilde A_{\bar{z}})^a+(D_{\bar{z}}\delta \tilde A_z)^a -(\tilde D_z\delta A_{\bar{z}})^a+(\tilde D_{\bar{z}}\delta A_z)^a }\\
=& -\frac{ik}{4\pi} \int \limits_{\Sigma} \varepsilon^a [-(D_z \delta A_{\bar{z}})^a - (D_z \delta E_{\bar{z}})^a + (D_{\bar{z}} \delta A_z)^a + (D_{\bar{z}} \delta E_z)^a\\
& \ \ \ \ \ \ \ \ \ \ \ \ \ \ -(D_z \delta A_{\bar{z}})^a - [E_z, \delta A_{\bar{z}}]^a +(D_{\bar{z}} \delta A_z)^a +[E_{\bar{z}},\delta A_z]^a]\\
=& -\delta \pa{\frac{ik}{4\pi} \int \limits_{\Sigma}\varepsilon^a \pb{2F^a_{z\bar{z}}+(D_z E_{\bar{z}})^a-(D_{\bar{z}}E_z)^a}}.
\es
}
\vskip0.35cm\noindent
Then using \eqref{eq:intcont2}, we can write the Gauss' law operator for TMYM theory as
\al{
G^a=\frac{ik}{4\pi} \pb{2F^a_{z\bar{z}}+(D_z E_{\bar{z}})^a-(D_{\bar{z}}E_z)^a}.
}


\section{The Wave-Functional}

We choose the K\"ahler polarization that makes $\psi$ only $A_{\bar{z}}$ and $\tilde{A}_{\bar{z}}$ dependent. The pre-quantum wave-functional is $\Phi[A_z,A_{\bar{z}},\tilde A_z, \tilde A_{\bar{z}}]=e^{-\frac{1}{2}K}\psi[A_{\bar{z}},\tilde A_{\bar{z}}]$, where $K$  is the K\"ahler potential given by $K=\frac{k}{4\pi}\int \limits_{\Sigma} (\tilde A^a_{\bar{z}} A^a_z+ A^a_{\bar{z}} \tilde A^a_z )$. Now, we can write
\al{
\label{eq:deltatmym}
A^a_z\psi=\frac{4\pi}{k} \frac{\delta}{\delta \tilde A^a_{\bar{z}}} \psi\ ~\text{and}\ ~\tilde A^a_z\psi=\frac{4\pi}{k} \frac{\delta}{\delta A^a_{\bar{z}}}\psi .
}
We make an infinitesimal gauge transformation on $\psi$ as
\al{
\label{eq:infgauge}
\delta_\varepsilon \psi[A_{\bar{z}},\tilde{A}_{\bar{z}}]=\int d^2z\ \pa{ \delta_\varepsilon A^a_{\bar{z}} \frac{\delta\psi}{\delta A^a_{\bar{z}}}  +\ \delta_\varepsilon \tilde{A}^a_{\bar{z}} \frac{\delta\psi}{\delta \tilde{A}^a_{\bar{z}}}}.
}
Using \eqr{eq:deltatmym} and $\delta A^a_{\bar{z}}=D_{\bar{z}}\varepsilon^a$, $\delta \tilde{A}^a_{\bar{z}}=\tilde{D}_{\bar{z}}\varepsilon^a$, we get
\vspace{.20cm}
\al{
\bs
\delta_\varepsilon \psi=& \int d^2z\ \varepsilon^a \pa{\tilde{D}_{\bar{z}}\frac{\delta}{\delta \tilde{A}^a_{\bar{z}}} + D_{\bar{z}}\frac{\delta}{\delta A^a_{\bar{z}}} } \psi\\
=&\frac{k}{4\pi} \int d^2z\ \varepsilon^a \pa { \partial_z \tilde{A}^a_{\bar{z}}+ \partial_z A^a_{\bar{z}} -2F_{z\bar{z}}- D_z E_{\bar{z}}+D_{\bar{z}}E_z  }\psi
\es
}
\vskip0.35cm\noindent
The generator of infinitesimal gauge transformations for this theory is
\al{
\label{eq:gausstmym}
G^a=\frac{ik}{4\pi}(2F_{z\bar{z}} + D_z E_{\bar{z}} - D_{\bar{z}}E_z)
}
while $G^a=\frac{ik}{2\pi}F_{z\bar{z}}$ being the generator for the pure CS theory as $E$-fields go to zero $(m\rightarrow\infty)$. After applying the Gauss' law $G^a\psi=0$, $\delta_\varepsilon \psi$ becomes
\al{
\label{eq:infg2}
\delta_\varepsilon \psi= \frac{k}{4\pi} \int d^2z\ \varepsilon^a \pa{\partial_{\bar{z}} \tilde A^a_z+\partial_{\bar{z}}  A^a_z} \psi.
}
This equation is similar to \eqr{eq:infg}. As they transform identically, $\tilde A$ can be parametrized the same way as $A$, using a different SL(N,$\mathbb{C}$) matrix $\tilde U$,
\al{
\tilde A_{\bar z}=-\partial_{\bar z}\tilde U \tilde U^{-1}~\text{and}~\tilde A_z=\tilde U^{{\dagger}{-1}}\partial_z \tilde U^{\dagger}.
}
The solution for the condition \eqr{eq:infg2} is
\al{
\label{eq:wftmym}
\psi[A_{\bar{z}},\tilde{A}_{\bar{z}}]=exp\pb{\frac{k}{2}\big(S_{WZW}(\tilde U)+S_{WZW}( U)\big)}\chi
}
where $\chi$ is the gauge invariant part of $\psi$ that is required to satisfy the Schr\"odinger's equation. Notice that \eqref{eq:wftmym}
reduces to Chern-Simons wave-functional \eqr{eq:cswf} as expected, when topological mass approaches infinity, which is equivalent to dropping the tilde symbol. $\chi$ should be equal to one in this limit.

To understand the role of the new matrix $\tilde{U}$, we can relate it to $U$ by rewriting \eqr{eq:Az} as
\al{
\label{eq:tilde}
\partial_{\bar z}\tilde U \tilde U^{-1}=\partial_{\bar z} U  U^{-1} + \frac{i}{2m}F^{0z}.
}
It turns out that $\tilde{U}$ is well behaved and solvable. Using the assumption $\tilde U=UM$, we can solve \eqref{eq:tilde} for $M$, viz;
\al{
\label{eq:M}
M(z,\bar{z})=\mc{P}exp \pa{ \frac{i}{2m} \int_0^{\bar{z}} \mc{F}^{0w}d\bar{w} }.
}
Here $\mc{F}^{0z}=U^{-1}F^{0z}U$ and it is gauge invariant. With this new gauge invariant matrix $M$, the electric field components can be written as
\al{
\label{eq:EU}
E_z=U^{\dagger -1}M^{\dagger -1}\partial_z M^\dagger U^\dagger \ ~\text{and}~\ E_{\bar{z}}=-U\partial_{\bar{z}} M M^{-1} U^{-1}.
}


\subsection{The Schr\"odinger's Equation}\label{sec:hamilt}

In the temporal gauge, the Hamiltonian gets no contribution from the Chern-Simons term. With no charges present, using $\alpha=\frac{4\pi}{k}$, $B=\frac{ik}{2\pi}F^{z\bar{z}}$ and  Euclidean metric, the TMYM Hamiltonian is given by
\al{
\mc{H}=\frac{m}{2\alpha}(E^a_{\bar{z}} E^a_z + E^a_z E^a_{\bar{z}})+\frac{\alpha}{m} B^a B^a.
}
Using \eqref{eq:deltatmym} we get
\vspace{0.20cm}
\al{
\bs
E^a_z(x)E^b_{\bar{z}}(x')\psi =&\ \alpha \pa{\frac{\delta}{\delta A^a_{\bar{z}}(x)} - \frac{\delta}{\delta \tilde{A}^a_{\bar{z}}(x)}}\big(\tilde A^b_{\bar{z}}(x')-A^b_{\bar{z}}(x')\big)\psi, 
\es
}
\vskip0.35cm\noindent
which gives the commutator
\al{
\label{eq:ecom}
[E^a_z(x),E^b_{\bar{z}}(x')]=-2 \alpha\ \delta^{ab}\delta^{(2)}(x-x').
}
Here, $E^a_z$ can be interpreted as an annihilation operator and $E^b_{\bar{z}}$ as a creation operator \cite{Grignani1997360}.
To get rid of the infinity, the Hamiltonian has to be normal ordered as
\al{
\mc{H}=\frac{m}{\alpha}E^a_{\bar{z}} E^a_z +\frac{\alpha}{m} B^a B^a.
}
To simplify the notation we write $\psi=\phi \chi$, where
\al{
\phi=exp\pb{\frac{k}{2}\big(S_{WZW}(\tilde U)+S_{WZW}( U)\big)}.
}
Derivatives of $\phi$ give the gauge field defined in \eqref{eq:scriptA} and its tilde version\cite{Nair:2005iw}, as
\al{
\label{eq:phi}
\tilde A^a_z\phi=\frac{4\pi}{k}\frac{\delta \phi}{\delta A^a_{\bar{z}}}=\mc{A}^a_z \phi\  ~\text{and}~\  A^a_z\phi=\frac{4\pi}{k}\frac{\delta \phi}{\delta \tilde{A}^a_{\bar{z}}}=\tilde{\mc{A}}^a_z \phi.
}
The holomorphic component of the $E$-field acting on $\psi$ is
\al{
E^a_z\psi=&\frac{4\pi}{k}\pa{\frac{\delta \phi}{\delta A^a_{\bar{z}}}-\frac{\delta \phi}{\delta \tilde{A}^a_{\bar{z}}}}\chi+\frac{4\pi}{k}\pa{\frac{\delta \chi}{\delta A^a_{\bar{z}}}-\frac{\delta \chi}{\delta \tilde{A}^a_{\bar{z}}}}\phi.
}
With defining $\mc{E}_z=\tilde{\mc{A}}_z - \mc{A}_z=-U\partial_z M M^{-1} U^{-1}$, we can write
\al{
\label{eq:EE}
E^a_z\psi=-\mc{E}^a_z\psi+\frac{4\pi}{k}\pa{\frac{\delta \chi}{\delta A^a_{\bar{z}}}-\frac{\delta \chi}{\delta \tilde{A}^a_{\bar{z}}}}\phi.
}
The magnetic field acting on $\psi$ is
\vspace{0.20cm}
\al{
\label{eq:B}
\bs
F^a_{z\bar{z}}\psi=&(\partial_z A^a_{\bar{z}}-D_{\bar{z}}A^a_z)\psi\\
=&D_{\bar{z}}(\mc{A}^a_z-A^a_z)\psi\\
=&D_{\bar{z}}\pa{-\mc{E}^a_z\psi-\frac{4\pi}{k}\frac{\delta \chi}{\delta \tilde{A}^a_{\bar{z}}}\phi}.
\es
}
\vskip0.35cm\noindent
The vacuum wave-functional is given by $\mc{H}\psi_0=0$, or
\al{
\label{eq:Hamilt}
E^a_{\bar{z}}E^a_z \psi_0 +\frac{1}{64m^2} F^a_{z\bar{z}} F^a_{z\bar{z}} \psi_0 =0.
}
The first term in \eqref{eq:Hamilt} is second order in $1/m$, while the Gauss' law forces the second term to be fourth order. We will continue our analysis with finite large values of $m$ where the potential energy term is negligible. Then, $E^a_z$ annihilates the vacuum as 
\al{
E^a_z\psi_0=0.
}
This condition is solved by writing
\vspace{0.20cm}
\al{
\label{eq:chi0}
\bs
\chi_0^{ }=&exp \pa{-\frac{k}{8\pi}\int \limits_{\Sigma} (\tilde{A}^a_{\bar{z}}-A^a_{\bar{z}})\mc{E}^a_z}=exp \pa{-\frac{k}{8\pi}\int \limits_{\Sigma}E^a_{\bar{z}}\mc{E}^a_z}.
\es
}
\vskip0.35cm\noindent
This solution is gauge invariant as required. In terms of $SL(N,\mathbb{C})$ matrices, it can be written as a function of only the gauge invariant matrix $M$, which was defined in \eqref{eq:M},
\al{
\chi_0^{ }=exp \pa{\frac{k}{4\pi}\int \limits_{\Sigma}Tr (E_{\bar{z}}\mc{E}_z)}=exp \pa{-\frac{k}{4\pi}\int \limits_{\Sigma}Tr (\partial_{\bar{z}}MM^{-1}\partial_zMM^{-1})}.
}
\vskip0.35cm\noindent

\section{The Measure}\label{sec:tmymmeasure}

Using the symplectic two-form \eqref{eq:omegatmym} we write the metric
\vspace{0.20cm}
\al{
\bs
ds^2_{\mathfrak{A}}=&-4\int Tr(\delta \tilde{A}_{\bar{z}} \delta A_z+\delta A_{\bar{z}}\delta \tilde{A}_z)\\
=&\ 4 \int Tr[\tilde{D}_{\bar{z}}(\delta \tilde{U} \tilde{U}^{-1})D_z(U^{\dagger -1}\delta U^{\dagger})+D_{\bar{z}}(\delta U U^{-1})\tilde{D}_z(\tilde{U}^{\dagger -1}\delta \tilde{U}^{\dagger})].
\es
}
\vskip0.35cm\noindent
Similar to the analysis in \ref{sec:csmeasure}, we find that the gauge invariant measure for this case to be
\al{
\label{eq:meas}
d\mu(\mathfrak{A})=det(\tilde{D}_{\bar{z}}D_z) det(D_{\bar{z}}\tilde{D}_z)d\mu(\tilde{U}^{\dagger}U)d\mu(U^{\dagger}\tilde{U}),
}
where for a certain choice of local counter terms ($\int Tr(\tilde A_{\bar{z}}A_z+\tilde A_z A_{\bar{z}})$),
\al{
det(\tilde{D}_{\bar{z}}D_z) det(D_{\bar{z}}\tilde{D}_z)=constant \times e^{2c_A\big(S_{WZW}(\tilde{U}^{\dagger}U)+S_{WZW}(U^{\dagger}\tilde{U})\big)}.
}
To simplify the notation we define a new matrix $N=\tilde{U}^{\dagger}U$. Since $\tilde{U}$ transforms like $U$, $N$ is gauge invariant. With this definition the measure becomes
\al{
\label{eq:measure}
d\mu(\mathfrak{A})=constant \times e^{2c_A\big(S_{WZW}(N)+S_{WZW}(N^{\dagger})\big)} d\mu(N)d\mu(N^{\dagger}).
}
To find the inner product, using the Polyakov-Wiegmann identity we write
\al{
e^{-K_{TMYM}}\psi_{TMYM}^*\psi^{ }_{TMYM}=e^{\frac{k}{2}\big(S_{WZW}(N)+S_{WZW}(N^{\dagger})\big)}\chi_0^*\chi^{ }_0
}
and from \eqref{eq:chi0} $\chi_0^*\chi^{ }_0$ (for large $m$) is 
\al{
\chi_0^*\chi^{ }_0=exp \pa{-\frac{k}{8\pi}\int \limits_{\Sigma}(E^a_z \mc{E}^a_{\bar{z}}+E^a_{\bar{z}}\mc{E}^a_z )}.
}
Then the inner product for the vacuum state is
\al{
\langle \psi_0|\psi_0\rangle=\int d\mu(N)d\mu(N^{\dagger})\ e^{(2c_A+\frac{k}{2})\big(S_{WZW}(N)+S_{WZW}(N^{\dagger})\big)}e^{-\frac{k}{8\pi}\int(E^a_z \mc{E}^a_{\bar{z}}+E^a_{\bar{z}}\mc{E}^a_z )}.
}
Since $\chi_0^*\chi^{ }_0=1+\mc{O}(1/m^2)$, we can neglect the second and higher order contributions at large scales compared to $1/m$, which leads to an almost topological theory in the near CS limit as
\al{
\label{eq:psipsi}
\langle \psi_0|\psi_0\rangle_{TMYM_k}{\approx} \int d\mu(N)d\mu(N^{\dagger})\ e^{(2c_A+\frac{k}{2})\big(S_{WZW}(N)+S_{WZW}(N^{\dagger})\big)}=\langle\psi|\psi\rangle_{CS_{k/2}}\langle\psi|\psi\rangle_{CS_{k/2}}.
}
Here the label ${TMYM_k}$ means that the inner product is calculated for TMYM theory with CS level number $k$. Similarly, $CS_{k/2}$ means the inner product is calculated for pure CS theory with level $k/2$. On the pure CS side, it is important to make this replacement of the level number to make the equivalence work. These two half-CS theories are not separately gauge invariant for odd values of $k$, but the sum of the two is. Each piece transforms as $\frac{1}{2}S_{CS}(A^g)\rightarrow\frac{1}{2}S_{CS}(A)+\pi k \omega(g)$ where $\omega(g)$ is the winding number. Then, the sum of the two will bring an extra $2\pi k \omega(g)$ that will not change the value of the path integral, even for odd values of $k$. In other words, any integer value of $k$ is sufficient to make the left hand side of \eqref{eq:psipsi} gauge invariant. But if one wants to make the two split CS parts separately gauge invariant on the right hand side of \eqref{eq:psipsi}, even values of $k$ must be chosen. 

This equivalence at large distances $(d>1/m)$ comes from the fact that the phase space of TMYM theory is a direct sum of two CS-like phase spaces. Thus, the classical equivalence discussed in refs. \citen{Lemes:1997vx, Lemes:1998md, Quadri:2002ni} does not work at the quantum level as it is, because of different phase space dimensionality of two theories.

In \eqref{eq:meas}, gauge invariance can be obtained in a different way by choosing other counter terms.
Choosing $\int Tr(A_{\bar{z}}A_z+\tilde A_z \tilde A_{\bar{z}})$ leads to
\al{
d\mu(\mathfrak{A})=constant \times e^{2c_A\big(S_{WZW}(H)+S_{WZW}(\tilde H)\big)} d\mu(H)d\mu(\tilde H).
}
With this option $e^{-K}\psi^*\psi$ part differs by $e^{-\int E^2}=1+\mc{O}(1/m^2)$ and the inner product can still be written in the form
\al{
\langle \psi_0|\psi_0\rangle_{TMYM_k}{\approx} \int d\mu(H)d\mu(\tilde H)\ e^{(2c_A+\frac{k}{2})\big(S_{WZW}(H)+S_{WZW}(\tilde H)\big)}=\langle\psi|\psi\rangle_{CS_{k/2}}\langle\psi|\psi\rangle_{CS_{k/2}}.
}
Thus, the Chern-Simons splitting can still be observed in the near Chern-Simons limit. Just like $N$ and $N^\dagger$, $H$ and $\tilde H$ are elements of $SL(N,\mathbb{C})/SU(N)$. However, the $N$, $N^\dagger$ seems to be a more \emph{natural} choice than $H$, $\tilde H$ because tilde and non-tilde variables are mixed in \eqref{eq:omegatmym}.

As we have seen in \autoref{sec:TMG}, the gravitational analogue of TMYM theory has a very similar CS splitting in the near CS limit, as
\al{
S[e]\approx\frac{1}{2\mu}S_{CS}\big[A^+[e]\big]+\frac{1}{2\mu}S_{CS}\big[A^-[e]\big].
}
This is analogous to what we have observed for TMYM theory at large distances, as we have predicted in \autoref{sec:TMG}.

\section{Wilson Loops in Topologically Massive Yang-Mills Theory}\label{sec:wilson}

With the new gauge field $\tilde{A}$, we can define a new loop operator
\al{
\label{eq:tildewlsn}
T_R(C)=Tr_R\ \mc{P}\ e^{-\oint \limits_c \tilde{A}_\mu dx^\mu}.
}
Just as the traditional Wilson loop, this operator is gauge invariant and it is an observable of the theory. To make a physical interpretation of this loop, we will check to see if it satisfies a 't Hooft-like algebra with the Wilson loop. To simplify the calculation, we will look at the abelian case with the following loops that live on $\Sigma$
\al{
W(C)=e^{i\oint \limits_c (A_z dz+A_{\bar{z}} d\bar{z})}\ ~ \text{and}\ ~ T(C)=e^{i\oint \limits_c (\tilde{A}_z dz+\tilde{A}_{\bar{z}} d\bar{z})}.
}
For the abelian case, the canonical relations differ from \eqref{eq:deltatmym} by a factor of 2. Then two loop operators satisfy a 't Hooft-like algebra
\al{
\label{eq:thooft}
T(C_1)W(C_2)=e^{\frac{2\pi i}{k}l(C_1,C_2)}W(C_2)T(C_1),
}
where $l(C_1,C_2)$ is the intersection number of $C_1$ and $C_2$ , which can only take values $0,\pm 1$. We cannot get a Dirac-like quantization condition since $k$ appears in the denominator and we want it to be a large integer to make the skein relations work. Therefore, the only option to make these operators commute is to have $l(C_1,C_2)=0$ thus, two loops cannot share a point. Equation \eqref{eq:thooft} lets us to interpret $T(C)$ as a 't Hooft-like loop for TMYM theory.

Working with the holomorphic polarization leads to the same problem we had in the pure Chern-Simons case. $A_z$ and $\tilde{A}_z$ are derivatives with respect to $\tilde{A}_{\bar{z}}$ and $A_{\bar{z}}$. This makes the path ordered exponential very complicated. To avoid this problem, we use a similar technique: Instead of using the traditional Wilson loop, we will calculate the expectation value of two loop operators that we define by $Tr\ U(x,x,C)$ and $Tr\ \tilde{U}(x,x,C)$ or
\al{
\mc{W}_R(C)=Tr_R\ \mc{P}\ e^{-\oint \limits_c (\mc{A}_zdz+A_{\bar{z}}d\bar{z})}\ ~
\text{and}~ \ 
\mc{T}_R(C)=Tr_R\ \mc{P}\ e^{-\oint \limits_c (\tilde{\mc{A}}_zdz+\tilde{A}_{\bar{z}}d\bar{z})}.
}
Once again these can be written in terms of WZW currents $-\partial_{\bar{z}}NN^{-1}$,  $-\partial_zNN^{-1}$,  $-\partial_{\bar{z}}N^{\dagger}N^{\dagger-1}$ and $-\partial_zN^{\dagger}N^{\dagger-1}$ as
\al{
\mc{W}_R(C)=Tr_R\ \mc{P}\ e^{\ \oint \limits_c (\partial_zNN^{-1}dz+\partial_{\bar{z}}NN^{-1}d\bar{z})}\ ~
\text{and}~ \ 
\mc{T}_R(C)=Tr_R\ \mc{P}\ e^{\ \oint \limits_c (\partial_zN^{\dagger}N^{\dagger-1}dz+\partial_{\bar{z}}N^{\dagger}N^{\dagger-1}d\bar{z})}.
}
These WZW currents are $SL(N,\mathbb{C})$ transformed $A$ and $\tilde A$ fields, just as in \eqref{eq:AJ}.

There is an interesting expectation value that we can calculate;
\vspace{0.20cm}
\al{
\bs
\langle \mc{W}_{R_1}(C_1)\mc{T}_{R_2}(C_2) \rangle=&\int d\mu(\mathscr{A})\ \psi_0^*\mc{W}_{R_1}(C_1)\mc{T}_{R_2}(C_2)\psi_0\\
=&\int d\mu(N)d\mu(N^{\dagger})\ e^{(2c_A+\frac{k}{2})\big(S_{WZW}(N)+S_{WZW}(N^{\dagger})\big)}\ \mc{W}_{R_1}(C_1,N)\mc{T}_{R_2}(C_2,N^{\dagger})\\
&+\mc{O}(1/m^2).
\es
}
\vskip0.35cm\noindent
This leads to an equivalence between the observables of CS and TMYM theory in the near CS limit. $\mc{W}_R(C)$ being only $N$ dependent and $\mc{T}_R(C)$ being only $N^{\dagger}$ dependent lets us to write
\begin{subequations}
\label{eq:equivalence}
\al{
\label{eq:equivalence1}
\langle \mc{W}_R(C)\rangle_{TMYM_{2k}} = \langle \mc{W}_R(C)\rangle_{CS_{k}}+\mc{O}(1/m^2),
}
\al{
\label{eq:equivalence2}
\langle \mc{T}_R(C)\rangle_{TMYM_{2k}} = \langle \mc{W}_R(C)\rangle_{CS_{k}}+\mc{O}(1/m^2)
}
and
\al{
\label{eq:equivalence3}
\langle \mc{W}_{R_1}(C_1)\mc{T}_{R_2}(C_2)\rangle_{TMYM_{2k}} = \bigg(\langle \mc{W}_{R_1}(C_1)\rangle_{CS_{k}}\bigg)\bigg(\langle \mc{W}_{R_2}(C_2)\rangle_{CS_{k}}\bigg)+\mc{O}(1/m^2).
}
\end{subequations}
To generalize these for $n$ loops, we can write
\begin{subequations}
\label{eq:equivalencegnrl}
\al{
\label{eq:equivalence1gnrl}
\langle \mc{W}_{R_1}(C_1)\textellipsis&\mc{W}_{R_n}(C_n)\rangle_{TMYM_{2k}} = \langle \mc{W}_{R_1}(C_1)\textellipsis\mc{W}_{R_n}(C_n))\rangle_{CS_{k}}+\mc{O}(1/m^2),
}
\al{
\label{eq:equivalence2gnrl}
\langle \mc{T}_{R_1}(C_1)\textellipsis&\mc{T}_{R_n}(C_n)\rangle_{TMYM_{2k}} = \langle\mc{W}_{R_1}(C_1)\textellipsis\mc{W}_{R_n}(C_n))\rangle_{CS_{k}}+\mc{O}(1/m^2)
}
and for mixed $n$ Wilson-like and $m$ 't Hooft-like loops,
\vspace{0.20cm}
\al{
\label{eq:equivalence3gnrl}
\bs
\langle \mc{W}_{R_1}(C_1)\textellipsis&\mc{W}_{R_n}(C_n)\mc{T}_{R'_1}(C'_1)\textellipsis\mc{T}_{R'_m}(C'_m)\rangle_{TMYM_{2k}}\\
& = \bigg(\langle \mc{W}_{R_1}(C_1)\textellipsis\mc{W}_{R_n}(C_n)\rangle_{CS_{k}}\bigg)\bigg(\langle\mc{W}_{R'_1}(C'_1)\textellipsis\mc{W}_{R'_m}(C'_m)\rangle_{CS_{k}}\bigg)\\
&+\mc{O}(1/m^2).
\es
}
\vskip0.35cm\noindent
\end{subequations}

Although \eqref{eq:psipsi} is gauge invariant even for odd values of $k$ as we have explained before, writing WLEVs of TMYM theory in terms of WLEVs of CS theory requires the two split CS parts to be separately gauge invariant. For this reason, we have used even level numbers on the left hand side of \eqref{eq:equivalence} and \eqref{eq:equivalencegnrl}. Notice that this gauge invariance issue arose only because we wanted to arrive to an equivalence between the observables of TMYM and CS theories. Otherwise, $\langle  \mc{W}_{R_1}(C_1)\mc{T}_{R_2}(C_2)\rangle_{TMYM_k}$ is gauge invariant in its own right for all integer values of $k$.

It seems like this set of equivalences also work for the case where $\Sigma=S^1\times S^1$. On a torus, similar to \eqref{eq:toruspsi}, TMYM wave-functional becomes
\al{
\psi[A_{\bar{z}},\tilde{A}_{\bar{z}}]=exp\pb{\frac{k}{2}(S_{WZW}(\tilde V)+S_{WZW}( V))}\Upsilon(\tilde{a})\Upsilon(a)\chi.
}
On the TMYM side, one needs to integrate over $V,\tilde{V},a,\tilde{a}$ and on the Chern-Simons side only over $V$ and $a$. Although it requires a more careful analysis, \eqref{eq:equivalencegnrl} seem to work on a torus as well. In principle, there is no reason to expect it to not work on any orientable $\Sigma$.

\chapter{2+1D Pure Yang-Mills Theory At Large Distances}\label{ch:ym}

In \autoref{ch:tmym}, we have seen that the inner product of topologically massive Yang-Mills(TMYM) theory can be written as
\al{
\label{eq:tmyminnerproduct}
\langle \psi_0|\psi_0\rangle_{TMYM_k}=\langle\psi|\psi\rangle_{CS_{k/2}}\langle\psi|\psi\rangle_{CS_{k/2}} +\mc{O}(1/m^2),
}
which is analogous to the Chern-Simons(CS) splitting of topologically massive AdS gravity model, which can be seen in \eqref{eq:tmgcs+cssplitting}. In the pure Einstein-Hilbert limit($\mu\rightarrow\infty$), \eqref{eq:tmg2CS} becomes
\al{
\label{eq:YMlimit}
S[e]=\frac{1}{2}S_{CS}\big[A^-[e]\big]-\frac{1}{2}S_{CS}\big[A^+[e]\big].
}
As we have seen in \autoref{sec:TMG}, the corresponding gauge theory for this limit is the pure Yang-Mills(YM) theory. With this motivation, we will study the pure YM theory in this chapter, following ref. \citen{yildirim2}. We will look at large distances to see whether or not the analogy can be extended to this limit and a similar splitting can be observed.


\section{Yang-Mills Theory In 2+1 Dimensions}\label{sec:YM}

The YM action is given by
\al{
\label{eq:actionym}
S_{YM}=-\frac{k}{4\pi}\frac{1}{4m}\int \limits_{\Sigma\times[t_i,t_f]} {d^3x }\ Tr\  (F_{\mu\nu}F^{\mu\nu}).
} 
The constant $\frac{k}{4\pi}$ is normally not necessary but here it is inserted for future convenience. The equations of motion are given by $D_{\nu}F^{\mu\nu}=0$.

In the temporal gauge $A_0=0$ with complex coordinates, the conjugate momenta are given by
\al{
\Pi^{a z}=\frac{k}{8\pi m}F^{a0z}~\ \text{and}~\ \Pi^{a \bar{z}}= \frac{k}{8\pi m}  F^{a0 \bar{z}}.
}
With $E_{z}=\frac{i}{2m}F^{0\bar{z}}$ and $E_{\bar{z}}=-\frac{i}{2m}F^{0z}$, the momenta can be rewritten as
\al{
\Pi^{a z}=\frac{ik}{4\pi} E^a_{\bar{z}}~\ \text{and}~\ \Pi^{a \bar{z}}=-\frac{ik}{4\pi} E^a_z.
}
With these coordinates, the symplectic two form is given by
\al{
\label{eq:omegaym}
\Omega=\frac{ik}{4\pi}\int \limits_{\Sigma}(\delta E^a_{\bar{z}} \delta A^a_z+\delta A^a_{\bar{z}}\delta E^a_z).
}
Similar to the calculation in \autoref{sec:gausscs} and \autoref{sec:gausstmym}, it can be shown that the generator of infinitesimal gauge transformations is $G^a=\frac{ik}{4\pi}(D_z E_{\bar{z}} - D_{\bar{z}}E_z)^a$.

To show how CS splitting happens, we need to write $\Omega$ in two \emph{CS-like} parts. To do this, we will use the coordinates
\al{
\label{eq:tildehatdef}
\tilde A_i=A_i+E_i\ ~\text{and}~\ \hat A_i=A_i-E_i.
}
Since $E_i$ is gauge covariant, both $\tilde A_i$ and $\hat A_i$ transform like gauge fields.
Now, $\Omega$ can be written as
\al{
\label{eq:omegaym2}
\Omega=\frac{ik}{4\pi}\int \limits_{\Sigma}(\delta \tilde A^a_{\bar{z}} \delta A^a_z-\delta A^a_{\bar{z}}\delta \hat A^a_z).
}
The phase space is K\"ahler with the potential 
\al{
\label{eq:kahlerpot}
K=\frac{k}{4\pi}\int \limits_{\Sigma}(\tilde A^a_{\bar{z}} A^a_z-A^a_{\bar{z}}\hat A^a_z).
}

Similar to what we have done before, we will parametrize the gauge fields. Since $\tilde A_i$ and $\hat A_i$ transform like gauge fields, they can be parametrized like one. Using new $SL(N,\mathbb{C})$ matrices $\tilde U$ and $\hat U$, defined similar to \eqref{eq:U}, we can write
\al{
\label{eq:partilde}
\tilde A_{\bar z}=-\partial_{\bar z}\tilde U\tilde U^{-1}~\text{and}~\tilde A_z=\tilde U^{\dagger-1}\partial_z \tilde U^{\dagger},
}
and
\al{
\label{eq:parhat}
\hat A_{\bar z}=-\partial_{\bar z}\hat U\hat U^{-1}~\text{and}~\hat A_z=\hat U^{\dagger-1}\partial_z \hat U^{\dagger}.
}
Also, $\tilde{\mc{A}}$ and $\hat{\mc{A}}$ can be written similar to \eqref{eq:scriptA}. We can use \eqref{eq:tildehatdef} to relate the matrices $\tilde U$ and $\hat U$ with $U$. Assuming $\tilde U=UM$ and $\hat U=UN$, we can solve \eqref{eq:tildehatdef} with
\al{
\label{eq:MM}
M(z,\bar{z})=N^{-1}(z,\bar{z})=\mc{P}exp \pa{ \frac{i}{2m} \int_0^{\bar{z}} \mc{F}^{0w}d\bar{w} },
}
where $\mc{F}^{0z}=U^{-1}F^{0z}U$. It can be seen that $\mc{F}$ is gauge invariant, so is $M$. With these new matrices, we can write the $E$-field components as
\vspace{0.20cm}
\al{
\label{eq:EU}
\bs
E_z=&U^{\dagger -1}M^{\dagger -1}\partial_z M^\dagger U^\dagger =-U^{\dagger -1}N^{\dagger -1}\partial_z N^\dagger U^\dagger\\
E_{\bar{z}}=&-U\partial_{\bar{z}} M M^{-1} U^{-1}=U\partial_{\bar{z}} N N^{-1} U^{-1}.
\es
}
\vskip0.35cm\noindent
We can also write the $\mc{E}$-fields that are given by $\mc{E}_i=\tilde{\mc{A}}_i-\mc{A}_i=\mc{A}_i-\hat{\mc{A}}_i$, as
\vspace{0.20cm}
\al{
\label{eq:scriptEU}
\bs
\mc{E}_z=&-U\partial_z M M^{-1} U^{-1}=U\partial_z N N^{-1} U^{-1}\\
\mc{E}_{\bar{z}}=&U^{\dagger -1}M^{\dagger -1}\partial_{\bar{z}} M^\dagger U^\dagger =-U^{\dagger -1}N^{\dagger -1}\partial_{\bar{z}} N^\dagger U^\dagger.
\es
}
\vskip0.35cm

\section{The Wave-Functional}

Choosing the holomorphic polarization gives $\Phi[A_z, A_{\bar{z}},\hat A_z, \tilde A_{\bar{z}}]=e^{-\frac{1}{2}K}\psi[ A_{\bar{z}},\tilde A_{\bar{z}}]$, where $K$ is the K\"ahler potential given in \eqref{eq:kahlerpot}, $\Phi$ and $\psi$ are the pre-quantum and quantum wave-functionals. $\chi$ is a function of both $A_{\bar{z}}$ and $\tilde A_{\bar{z}}$ and since it is gauge invariant, it has to be a function of the difference of these variables, which is $E_{\bar{z}}$.

From \eqref{eq:omegaym2}, upon quantization we can write
\al{
\label{eq:deltaym}
A^a_z\psi=\frac{4\pi}{k} \frac{\delta\psi}{\delta \tilde A^a_{\bar{z}}} \ ~\text{and}\ ~\hat A^a_z\psi=-\frac{4\pi}{k} \frac{\delta\psi}{\delta A^a_{\bar{z}}}.
}
Now, we make an infinitesimal gauge transformation on $\psi$,
\al{
\label{eq:infgauge}
\delta_\epsilon \psi[A_{\bar{z}},\tilde{A}_{\bar{z}}]=\int d^2z\ \pa{ \delta_\epsilon A^a_{\bar{z}} \frac{\delta\psi}{\delta A^a_{\bar{z}}}  +\ \delta_\epsilon \tilde{A}^a_{\bar{z}} \frac{\delta\psi}{\delta \tilde{A}^a_{\bar{z}}}}.
}
Using \eqr{eq:deltaym}, $\delta A^a_{\bar{z}}=D_{\bar{z}}\epsilon^a$ and $\delta \tilde{A}^a_{\bar{z}}=\tilde{D}_{\bar{z}}\epsilon^a$, \eqr{eq:infgauge} can be rewritten as
\vspace{0.20cm}
\al{
\label{eq:infgauge2}
\bs
\delta_\epsilon \psi=& \int d^2z\ \epsilon^a \pa{ D_{\bar{z}}\frac{\delta}{\delta A^a_{\bar{z}}} + \tilde{D}_{\bar{z}}\frac{\delta}{\delta \tilde{A}^a_{\bar{z}}} } \psi\\
=&\frac{k}{4\pi} \int d^2z\ \epsilon^a \pa { \tilde D_{\bar{z}}A^a_z - D_{\bar{z}}\hat A^a_z }\psi.
\es
}
\vskip0.35cm\noindent
Using \eqref{eq:tildehatdef}, \eqref{eq:infgauge2} becomes
\al{
\delta_\epsilon \psi=&\frac{k}{4\pi} \int d^2z\ \epsilon^a \pa { \partial_z E^a_{\bar{z}} - D_z E^a_{\bar{z}} + D_{\bar{z}}E^a_z }\psi.
}
After applying the Gauss' law $G^a\psi=0$, we get
\vspace{0.20cm}
\al{
\label{eq:infgauge3}
\bs
\delta_\epsilon \psi=&\frac{k}{4\pi} \int d^2z\ \epsilon^a \pa { \partial_z E^a_{\bar{z}} }\psi\\
=&\frac{k}{4\pi} \int d^2z\ \epsilon^a \pa { \partial_z \tilde A^a_{\bar{z}}-\partial_z A^a_{\bar{z}} }\psi\\
=&\frac{k}{4\pi} \int d^2z\ \epsilon^a \pa {\partial_z  A^a_{\bar{z}}-\partial_z \hat A^a_{\bar{z}} }\psi.
\es
}
\vskip0.35cm\noindent
This condition is solved by $\psi=\phi\chi$ with
\al{
\label{eq:wfym1}
\phi[A_{\bar{z}},\tilde{A}_{\bar{z}}]=exp\pb{\frac{k}{2}\big(S_{WZW}(\tilde U)-S_{WZW}( U)\big)},
}
or equally
\al{
\label{eq:wfym2}
\phi[A_{\bar{z}},\tilde{A}_{\bar{z}}]=exp\pb{\frac{k}{2}\big(S_{WZW}(U)-S_{WZW}( \hat U)\big)}.
}
The equivalence of \eqref{eq:wfym1} and \eqref{eq:wfym2} can be shown by using the Polyakov-Wiegmann(PW) identity with $\tilde U=UM$ and $\hat U=UM^{-1}$.

Holomorphic components of gauge fields acting on $\phi$ gives the $\mc{A}$ and $\tilde{\mc{A}}$ fields, as
\al{
\label{eq:phi}
A^a_z\phi=\frac{4\pi}{k}\frac{\delta \phi}{\delta \tilde A^a_{\bar{z}}}=\mc{A}^a_z \phi\  ~\text{and}~\  \hat A^a_z\phi=-\frac{4\pi}{k}\frac{\delta \phi}{\delta A^a_{\bar{z}}}=\tilde{\mc{A}}^a_z \phi.
}


\subsection{The Schr\"odinger's Equation}\label{sec:schrodinger}

The YM Hamiltonian is given by
\al{
\mc{H}=T+V\ ~\text{with}~\ T=\frac{m}{\alpha}E^a_{\bar{z}} E^a_z\ ~\text{and}~\  V=\frac{\alpha}{m} B^a B^a,
}
where $\alpha=\frac{4\pi}{k}$, $B=\frac{ik}{2\pi}F^{z\bar{z}}$. We are interested in finding the vacuum wave-functional, which is given by $\mc{H}\psi_0=0$, or with using Euclidean metric,
\al{
\label{eq:Hamilt}
E^a_{\bar{z}}E^a_z \psi_0 +\frac{1}{64m^2} F^a_{z\bar{z}} F^a_{z\bar{z}} \psi_0 =0.
}
For both YM and TMYM theories, $\chi$ is typically in the form of $e^{-\frac{1}{m^2}\int F^2}$\cite{Karabali:1998yq,Karabali:1999ef,yildirim}(except for very small distances), since they have the same Hamiltonian in the temporal gauge. This exponential decay behavior cannot be polarization dependent and it comes from the existence of a mass gap. Thus, we expect it to be valid here as well. To check this assumption, we will first focus on the potential energy term. Using \eqref{eq:phi}, the B-field acting on $\psi$ is
\vspace{0.20cm}
\al{
\label{eq:B}
\bs
F^a_{z\bar{z}}\psi=&(\partial_z A^a_{\bar{z}}-D_{\bar{z}}A^a_z)\psi\\
=&D_{\bar{z}}(\mc{A}^a_z-A^a_z)\psi\\
=&D_{\bar{z}}\pa{-\mc{E}^a_z\psi-\frac{4\pi}{k}\frac{\delta \chi}{\delta \tilde{A}^a_{\bar{z}}}\phi},
\es
}
\vskip0.35cm\noindent
where $\mc{E}_z$ is given by $\tilde{\mc{A}}_z-\mc{A}_z$. If we consider only the potential energy term, the gauge invariant part of the vacuum wave-function is given by
\al{
\label{eq:chiB}
\chi=exp\pa{-\frac{k}{4\pi}\int \limits_{\Sigma}E^a_{\bar{z}}\mc{E}^a_z}=exp\pa{-\frac{k}{4\pi}\int \limits_{\Sigma}(\tilde A^a_{\bar{z}}-A^a_{\bar{z}})\mc{E}^a_z}=1+\mc{O}(1/m^2).
}
Now, to study the kinetic energy term we look at $E_z\psi$. Using \eqref{eq:phi}, we can write
\vspace{0.20cm}
\al{
\bs
E^a_z\psi=&(A^a_z-\hat A^a_z)\phi\chi\\
=&\mc{E}^a_z \psi + \frac{4\pi}{k}\phi \pa{\frac{\delta\chi}{\delta \tilde A^a_{\bar{z}}}+\frac{\delta\chi}{\delta A^a_{\bar{z}}}}.
\es
}
\vskip0.35cm\noindent
Since $\chi$ is gauge invariant, we can write $\chi=\chi[\tilde A^a_{\bar{z}},A^a_{\bar{z}}]=\chi[\tilde A^a_{\bar{z}}-A^a_{\bar{z}}]$. This leads to
\al{
\frac{\delta\chi}{\delta \tilde A^a_{\bar{z}}}=-\frac{\delta\chi}{\delta A^a_{\bar{z}}}=\frac{\delta\chi}{\delta E^a_{\bar{z}}}.
}
Then, we get
\al{
E^a_{\bar{z}}E^a_z\psi = E^a_{\bar{z}}\mc{E}^a_z\psi,
}
which has no contribution from $\chi$. Thus, as far as only the kinetic energy term is concerned, a constant $\chi$ is sufficient\cite{Karabali:1998yq}. If we neglect the potential energy term, the vacuum condition forces $\mc{E}^a_z$ to be zero. This means that the matrix $M$ has to be a holomorphic function of $\bar{z}$. 

Our goal is to study the large distance behavior of the theory by neglecting second and higher order terms in $1/m$. As we have done for TMYM theory, in \autoref{sec:hamilt}, neglecting the potential energy term is standard practice in these type of cases. In TMYM theory, Gauss' law forced the magnetic fields to be first order in $1/m$ but that is not the case here. But in the literature, magnetic field contribution still gets neglected with the following reasoning. Since the energy is low, charges move very slowly and do not create significant magnetic fields\cite{Grignani1997360}. When the potential term is not neglected, we expect the full solution for \eqref{eq:Hamilt} to be an interpolation between the kinetic energy eigenstate and potential energy eigenstate. Since neither of these states has a first order term in $1/m$, even if we do not neglect the potential term, there should not be any first order contribution. This result is consistent with ref. \citen{Karabali:1998yq}. Thus, for the scales that we study, $\chi=1$ can be taken safely (at least when the potential energy is neglected), just like in TMYM theory. 


\section{The Measure}

From \eqref{eq:kahlerpot}, we can write the metric for the space of gauge fields $\mathfrak{A}$ as
\vspace{0.20cm}
\al{
\bs
ds^2_{\mathfrak{A}}=&-4\int Tr(\delta \tilde{A}_{\bar{z}} \delta A_z-\delta A_{\bar{z}}\delta \hat{A}_z)\\
=&\ 4 \int Tr[\tilde{D}_{\bar{z}}(\delta \tilde{U} \tilde{U}^{-1})D_z(U^{\dagger -1}\delta U^{\dagger})-D_{\bar{z}}(\delta U U^{-1})\hat{D}_z(\hat{U}^{\dagger -1}\delta \hat{U}^{\dagger})].
\es
}
\vskip0.35cm\noindent
Similar to the analysis in \autoref{sec:tmymmeasure}, the gauge invariant measure is
\al{
\label{eq:meas}
d\mu(\mathfrak{A})=det(\tilde{D}_{\bar{z}}D_z) det(D_{\bar{z}}\hat{D}_z)d\mu(\hat{U}^{\dagger}U)d\mu(U^{\dagger}\tilde{U}).
}
For a certain choice of local counter terms ($\int Tr(\tilde A_{\bar{z}}A_z+\hat A_z A_{\bar{z}})$) we get
\al{
det(\tilde{D}_{\bar{z}}D_z) det(D_{\bar{z}}\hat{D}_z)=constant \times e^{2c_A\big(S_{WZW}(U^{\dagger}\tilde{U})+S_{WZW}(\hat{U}^{\dagger}U)\big)}.
}
To simplify the notation, we will continue with defining $H_1=U^{\dagger}\tilde{U}$ and $H_2=\hat{U}^{\dagger}U$. These two matrices are $SU(N)$ gauge invariant and belong to the coset $SL(N,\mathbb{C})/SU(N)$. Now, the gauge invariant measure can be written as
\al{
\label{eq:meas2}
d\mu(\mathfrak{A})=e^{2c_A\big(S_{WZW}(H_1)+S_{WZW}(H_2)\big)}d\mu(H_1)d\mu(H_2).
}


\section{Chern-Simons Splitting}

As we have shown in \autoref{sec:geoquan}, for holomorphic polarizations, the inner product is given by
\al{
\langle \psi|\psi \rangle=\int d\mu(\mathfrak{A})\ e^{-K}\ \psi^*\psi^{ },
}
where $K$ is the K\"ahler potential. Using \eqref{eq:kahlerpot}, \eqref{eq:wfym1} and \eqref{eq:wfym2} with PW identity gives
\al{
\label{eq:psipsiym}
e^{-K}\ \psi^*\psi^{ }=e^{\frac{k}{2}\big(S_{WZW}(H_1)-S_{WZW}(H_2)\big)}\chi^*\chi^{ }.
}
As we have seen in \autoref{sec:schrodinger},
\al{
\label{eq:chiorder}
\chi^*\chi^{ }=1+\mc{O}(1/m^2).
}
Now, using \eqref{eq:meas2}, \eqref{eq:psipsiym} and \eqref{eq:chiorder}, we can write the inner product as
\al{
\label{eq:YMinnerproduct}
\langle \psi_0|\psi_0 \rangle=\int d\mu(H_1)d\mu(H_2)\ e^{\pa{2c_A+\frac{k}{2}}S_{WZW}(H_1)+\pa{2c_A-\frac{k}{2}}S_{WZW}(H_2)}+\mc{O}(1/m^2).
}
Here, it can be seen that the YM inner product consists of two half-level CS parts with opposite signs, that cancel as $m\rightarrow\infty$, since $H_1=H_2=H$ in this limit. The YM inner product can be written as
\al{
\label{eq:YMsplit}
\langle \psi_0|\psi_0\rangle_{YM_k}=\langle \psi|\psi \rangle_{CS_{k/2}}\langle \psi|\psi \rangle_{CS_{-k/2}}+\mc{O}(1/m^2).
}
It appears that YM inner product splits into two CS terms at large distances, just as we predicted by studying its gravitational analogue \eqref{eq:YMlimit}.

\chapter{Conclusion}\label{ch:conc}

We have shown that due to the existence of a mass gap, topologically massive Yang-Mills(TMYM) theory in the \emph{near} Chern-Simons(CS) limit is an almost topological field theory that consists of two copies of CS, similar to the topologically massive AdS gravity model. One copy is associated with the matrix $N$ and the other with $N^{\dagger}$, each with half the level of the original CS term in the TMYM Lagrangian. Separately momentum and position Hilbert spaces of TMYM theory can be thought of CS Hilbert spaces with half the level. In the $m\rightarrow\infty$ limit, where $N=N^{\dagger}=H$, these two CS theories add up to give one CS with the original level number $k$, as
\al{
e^{\frac{k}{2}\big(S_{WZW}(N)+S_{WZW}(N^{\dagger})\big)} \overset{m\rightarrow\infty}\longrightarrow e^{kS_{WZW}(H)}.
}
Although the integrand behaves well as $m\rightarrow\infty$, this limit is delicate for the integral measure. Studying large values of $m$ does not cause any problem, but taking it to infinity reduces the phase space dimension from four to two; thus, a change in the integral measure becomes necessary. Except for this phase space reduction, dropping the tilde symbol gives the correct CS limit in our calculations. In this limit, the metric of the space of gauge potentials reduces as
\al{
\label{eq:reduction}
-4\int Tr(\delta \tilde{A}_{\bar{z}} \delta A_z+\delta A_{\bar{z}}\delta \tilde{A}_z)\ 
\overset{m\rightarrow\infty}\longrightarrow\ -8\int Tr(\delta A_{\bar{z}} \delta A_z).
}
For the left hand side of \eqref{eq:reduction}, the measure is given by \eqref{eq:measure} while for the right hand side, it is given by \eqref{eq:CSmeasure}. Thus, the measure needs to be replaced with \eqref{eq:CSmeasure} in the pure CS limit. Although this reduction occurs beautifully in the metric, the volume \eqref{eq:measure} does not automatically reduce to \eqref{eq:CSmeasure} in our notation. Not switching to the correct volume element results in duplicate integration over $H$, since in the pure CS limit $N=N^{\dagger}=H$. This comes from the fact that the phase space of TMYM theory consists of two CS phase spaces.

A different problem exists for the pure Yang-Mills(YM) limit $m\rightarrow0$. In this case, dimensionality of the phase space does not change, but since $E$-fields do not gauge transform like $\tilde{A}$ fields, our parametrization and measure do not work. But the main problem with studying the pure YM limit comes from not knowing the magnetic field contribution in the wave-functional, which becomes the dominant part in this limit. To get the magnetic field contribution, \eqref{eq:Hamilt} needs to be fully solved without using the strong coupling limit.

CS splitting does not seem to appear if one uses the real polarization that makes $\psi=\psi[A_z, A_{\bar{z}}]$. This can be seen clearly in refs. \citen{Asorey1993477,Grignani1997360,Karabali:1999ef}. These earlier works study TMYM theory in a perspective where YM theory is perturbed by a CS term, while our work is exactly the opposite. Ideally, all polarizations should lead to the same inner product; but in this case, different polarizations create different mathematical difficulties that force one to focus on certain scales. For TMYM theory, holomorphic polarization facilitates studies near the CS limit, while the real polarization is suitable for the near YM limit\cite{nair}. The difficulties that arise when using the real polarization can be summarized as follows. First, $A_z$ and $ A_{\bar{z}}$ do not commute in the pure CS limit; the wave-functional cannot depend on both, at very large distances. This makes it impossible to check the CS limit of the wave-functional. But in the holomorphic polarization TMYM wave-functional reduces to CS wave-functional nicely. Second, in the real polarization, the wave functional cannot depend on $E$-fields. But in the near CS limit, $E$-fields dominate the Hamiltonian while $B$-fields are negligible. This creates a difficulty in studying large distances. Since CS splitting occurs at a scale where the first order $E$-field contributions are important, it is not surprising to not see this feature in a polarization that cannot resolve this scale. Similarly, holomorphic polarization is not helpful in studying smaller scales where $B$-field contributions are important, since $\psi[A_{\bar{z}},\tilde A_{\bar{z}}]$ does not depend on $B$-fields. Thus, to study the near CS limit, holomorphic polarization should be chosen. Therefore, our results cannot be compared\cite{nair} with the results of refs. \citen{Asorey1993477,Grignani1997360,Karabali:1999ef}, since they focus on a different scale where CS splitting does not occur.

Usually in the literature, the constant $k/4\pi$ is not inserted in the YM term and this leads to a mass gap $\propto km$. Thus, a question may arise on whether or not taking $k$ to be a large integer causes any problem. In our calculations, it did not cause any problem in obtaining a topological theory at large distances. This indicates that having a large level number does not alter the existence of the mass gap in TMYM theory, although we do not provide a proof.

In \autoref{sec:wilson}, by writing \eqref{eq:equivalencegnrl} we showed that loop operator expectation values of CS and TMYM theories are related at large distances. This equivalence tells us that expectation values of both Wilson loops and 't Hooft loops in TMYM theory are equal to CS Wilson loop expectation values, up to a change in level number. A more interesting result is that the expectation value of the product of these loops in TMYM theory is equal to the product of Wilson loop expectation values in CS theory. These results show that not only in the pure CS limit but also in the near CS limit, the observables of TMYM theory are link invariants. Both Wilson loops and 't Hooft loops can separately form links that satisfy the skein relation \eqref{eq:skeinw}, but a mixed link of these loops does not, even though it is still a link invariant.

In \autoref{ch:ym}, we have shown that YM theory acts like a topological theory at certain limited scales, just like TMYM theory. When the distance is large enough but finite, YM theory splits into two CS terms with levels $k/2$ and $-k/2$ very similar to the splitting of topologically massive AdS gravity model. At very large distances, these two CS terms cancel to make YM theory trivial. Together with our calculation for TMYM theory, we have shown that both YM and TMYM theories exhibit CS splitting at large scales, just as predicted by their gravitational analogues. The methods that we have introduced in \autoref{sec:wilson} can be used to exploit this limited topological region to incorporate link invariants for pure YM theory as well.

\biblio{Prelude/bibliography.bib} 

\end{document}